\documentclass[12pt,twocolumn,apj]{aastex631}

\usepackage{changepage}
\usepackage{booktabs}
\usepackage{aas_macros}
\usepackage{epsfig}
\usepackage{amsmath}
\usepackage{amssymb}
\usepackage{comment}
\usepackage{leftidx}
\definecolor{MyGreen}{rgb}{0.0,0.6,0.3}
\definecolor{MyPurple}{rgb}{0.6,0,0.3}
\usepackage{fancyref}
\usepackage{hyperref}
\hypersetup{colorlinks=true,linkcolor=MyPurple,urlcolor=MyPurple,citecolor=MyGreen}
\usepackage{color,colortbl}
\usepackage{tensind}
\tensordelimiter{?}
\DeclareGraphicsExtensions{.bmp,.png,.jpg,.pdf}
\usepackage{verbatim}
\usepackage[normalem]{ulem}
\usepackage{orcidlink}
\usepackage{soul}
\usepackage{upgreek}
\def\beq{\begin{equation}}
\def\eeq{\end{equation}}
\def\ba{\begin{eqnarray}}
\def\ea{\end{eqnarray}}
\def\bal{\begin{align}}
\def\eal{\end{align}}
\def\bxi{{\mbox{\boldmath $\xi$}}}

\begin{document}

\title[triaxial pulsators] {Tidally distorted stars are triaxial pulsators}

\author{Jim Fuller\orcidlink{0000-0002-4544-0750}}
\email{jfuller@caltech.edu}
\affiliation{TAPIR, Mailcode 350-17, California Institute of Technology, Pasadena, CA 91125, USA}

\author{Saul Rappaport}
\affiliation{Department of Physics, Kavli Institute for Astrophysics and Space Research, M.I.T., Cambridge, MA 02139, USA}

\author{Rahul Jayaraman}
\affiliation{Department of Physics, Kavli Institute for Astrophysics and Space Research, M.I.T., Cambridge, MA 02139, USA}

\author{Don Kurtz}
\affiliation{Centre for Space Research, North-West University, Dr Albert Luthuli Drive, Mahikeng 2735, South Africa}
\affiliation{Jeremiah Horrocks Institute, University of Central Lancashire, Preston PR1 2HE, UK}
\author{Gerald Handler}

\affiliation{Nicolaus Copernicus Astronomical Center, Polish Academy of Sciences, ul. Bartycka 18, 00-716, Warsaw, Poland}

\begin{abstract}

Stars in close binaries are tidally distorted, and this has a strong effect on their pulsation modes. We compute the mode frequencies and geometries of tidally distorted stars using perturbation theory, accounting for the effects of the Coriolis force and the coupling between different azimuthal orders $m$ of a multiplet induced by the tidal distortion. For tidally coupled dipole pressure modes, the tidal coupling dominates over the Coriolis force and the resulting pulsations are ``triaxial", with each of the three modes in a multiplet ``tidally tilted" to be aligned with the one of the three principal axes of the star. The observed amplitudes and phases of the dipole modes aligned orthogonal to the spin axis are modulated throughout the orbit, producing doublets in the power spectrum that are spaced by exactly twice the orbital frequency. Quadrupole modes have similar but slightly more complex behavior. This amplitude modulation allows for mode identification which can potentially enable detailed asteroseismic analyses of tidally tilted pulsators. Pressure modes should exhibit this behavior in stellar binaries close enough to be tidally synchronized, while gravity modes should remain aligned with the star's spin axis. We discuss applications to various types of pulsating stars, and the relationship between tidal tilting of pulsations and the ``single-sided" pulsations sometimes observed in very tidally distorted stars.

\end{abstract}



\section{Introduction}


Most asteroseismic calculations assume that stars are spherical, simplifying calculations of their oscillation modes. Pulsations of rotating stars are subject to Coriolis and centrifugal forces, which break the spherical symmetry and define the rotation axis as the symmetry axis of the pulsations. The majority of models thus assume that each stellar oscillation mode has an angular pattern of a spherical harmonic $Y_{lm}(\theta,\phi)$ defined relative to the star's rotation axis. Exceptions to this are the rapidly oscillating Ap stars (\citealt{1982MNRAS.200..807K}) where the pulsation symmetry axis closely aligns with the oblique magnetic axis. See \citet{2021FrASS...8...31H} and \citet[][section 2.5.12]{2022ARA&A..60...31K} for introductions and reviews of these stars.

Stars in close binaries, however, are distorted by the tidal force of the companion, which is symmetric about the tidal axis (i.e., the line connecting the two stars). Many such stars are also tidally synchronized so that the rotation rate is equal to the orbital frequency. In this case, centrifugal and tidal distortion are of similar magnitude, but symmetric about orthogonal axes. Such stars are hence triaxial ellipsoids, with their shortest axis (the $z$-axis) along the rotation/orbital axis, their longest axis along the tidal axis (the $x$-axis), and their intermediate axis perpendicular to these two (the $y$-axis). The goal of this paper is to compute the pulsation geometries of such triaxial stars.

Initial work investigating pulsations of tidally distorted stars (e.g., \citealt{chandrasekhar1963,chandrasekhar1963b,tassoul1967,denis1972} largely focused on mode stability, often investigating the fundamental modes of constant density or polytropic stellar models. Subsequent work investigated the frequency perturbations of modes due to small tidal distortion \cite{saio:81,martens1982,smeyers1983}.

Perhaps the most crucial effect of tides is to change the pulsation geometry such that the modes are no longer spherical harmonics with respect to the rotation axis. \cite{reyniers2003,reyniers2003b} computed the pulsation geometry of tidally distorted stars, showing that tidal distortion couples modes of different azimuthal order $m$, such that the eigenmodes are superpositions of modes of differing $m$ (see also \citealt{martens1986}). They found that modes have the tidal axis as their symmetry axis, and so the $(2\ell+1)$ modes of a multiplet with angular degree $\ell$ are split into $(\ell+1)$ different frequencies in the rest frame of the star. However, those works did not include the centrifugal force which disrupts the symmetry around the tidal axis by making the stars triaxial.

\cite{fuller:20} used a similar method to compute the pulsations of heavily tidally distorted stars. They accounted for tidal coupling between modes of different angular degree $\ell$ and radial order $n_{\rm pg}$, which can cause a mode's surface flux perturbations to be confined primarily to one side of the star. However, they assumed that the modes are aligned with the tidal axis, i.e., each mode has a single azimuthal number $m$ that defines the mode pattern number around the tidal axis. This assumption is not justified when one considers the Coriolis and centrifugal forces that break the symmetry around the tidal axis and creates triaxially distorted stars.

If a mode produces a spherical harmonic pattern aligned with the tidal axis rather than the rotation axis, its amplitude and phase will be observed to vary over the course of the orbit. Such a ``tidally tilted" pulsation is split into multiple components in the observed power spectrum, spaced by integer multiples of the orbital frequency. This modulation is straightforward to predict \citep{reed:05,balona2018}, causing dipole modes to typically produce a triplet spaced by the orbital frequency in the observed power spectrum. However, we will demonstrate that this is not the correct geometry of tidally tilted oscillations, altering expectations for the observed power spectra.

Observations of pulsating and tidally distorted stars have increased enormously due to data from the \textit{Kepler} and \textit{TESS} satellites.  \cite{handler:20} and \cite{kurtz:20} found tidally distorted $\delta$\,Scuti stars whose pulsations are confined to the side of the star facing its companion. \cite{rappaport:21} found a $\delta$\,Scuti star with a pulsation confined to the tidal equator. \cite{kahraman:22} examined a $\delta$\,Scuti pulsator in a close binary with several doublets spaced by twice the orbital frequency, indicative of tidally tilted pulsations. \cite{jayaraman:22} found an sdB pulsator in a close binary that exhibits a rich diversity of tidally tilted pulsations, showing that most of these could be identified as primarily $\ell=1$ or $\ell=2$ tidally tilted modes. The density of low-$\ell$ modes also suggested the presence of mixed modes, which indicated the star had recently finished helium core-burning. Other pulsating stars which appear to exhibit tidally perturbed p~modes include KPD 1930+2752 \citep{reed:11}, U Gru \citep{Bowman2019,johnston:23}, KIC 4142768 \citep{balona2018}, and KIC 9851944 \citep{jennings:24}. Tidally perturbed g~modes have also been observed in several stars \citep{vanreeth:22,vanreeth:23}.

Most recently, \cite{zhang:24} found a $\delta$\,Scuti pulsator with many tidally tilted $\ell=1$ modes. Each such mode forms a nearly equal-amplitude doublet in the power spectrum spaced by 2$\times$ the orbital frequency. Importantly, they showed that the observed pulsation amplitude and phase modulation could not arise from spherical harmonics aligned with the tidal axis. Instead, they showed that the data could be explained by ``triaxial" pulsations, each symmetric about one of the star's three ellipsoidal axes, and they presented a theoretical calculation demonstrating why this is expected in tidally distorted stars. \cite{jayaraman:24} also found a $\delta$\,Scuti pulsator with a large number of predominantly triaxial modes.

In this work, we expand upon the simple calculation of \cite{zhang:24} to compute the mode frequencies and mode geometry of a tidally distorted star, including the effects of tidal distortion, centrifugal distortion, and the Coriolis force. We show that the pulsations of triaxial stars are themselves triaxial, as proposed by \cite{zhang:24}. For the three $\ell=1$ modes of a given radial order, the modes are each aligned with one of the three principal axes of the triaxial star, creating three unique mode frequencies in the frame of the star. In the observer's frame, however, the changing viewing angle during the orbit causes the observed amplitudes and phases to vary, causing each mode to appear as a doublet or singlet in the power spectrum.
We perform the same calculations for the star's five quadrupole ($\ell=2$) pulsations, finding the angular pattern and frequency perturbation for each. We provide simple formulae to accurately estimate the tidal frequency perturbations in Section \ref{sec:approx}, we apply our method to $\delta$~Scuti pulsators in Section \ref{sec:application}, and we discuss when we expect to observe triaxial and tidally tilted pulsations in Section \ref{sec:discussion}.

\section{Tidal Coupling}

Our goal is to compute the pulsation mode geometry for a tidally distorted star using linear perturbation theory. Our analysis is is similar to that of \cite{fuller:20}, but they assumed the modes have a symmetry axis along the tidal axis (the line connecting the two stars). In this work, we self-consistently compute the pulsation geometry, but we only consider perturbations to isolated multiplets. In other words, for a mode multiplet with angular order $\ell$ and radial order $n_{\rm pg}$, we compute coupling between the different azimuthal numbers $m$ within the multiplet, but we do not account for coupling with different angular orders $\ell$ or radial orders $n_{\rm pg}$.

In this paper, we will assume that the pulsating star is in a circular orbit, and that its spin is synchronized and aligned with the companion. We decompose the tidal potential into terms with spherical harmonic dependence $Y_{\ell_t,m_t}(\theta,\phi)$, and we only consider the dominant $\ell_t=2$ component of the tidal distortion, since the $\ell_t=3$ component is smaller by a factor of $R/a$, where $R$ is the stellar radius and $a$ is the semi-major axis. This approximation will not be appropriate for stars nearly filling their Roche lobes, where $\ell_t \geq 3$ components should be included.

Following the framework outlined in \cite{dahlen:98} and \cite{fuller:14}, stellar oscillation modes are described by solutions to the generalized eigenvalue problem
\beq
\label{eq:mat1}
\begin{bmatrix} 0 & \mathcal{V} \\ \mathcal{V} & 2 \mathcal{W} \end{bmatrix} {\bf z} = \omega \begin{bmatrix} \mathcal{V} & 0 \\ 0 & \mathcal{T} \end{bmatrix} {\bf z},
\eeq
where 
\beq
\label{eq:z}
{\bf z} = \left[ \begin{array}{c} {\bxi} \\ \omega {\bxi} \end{array} \right] \, ,
\eeq
and ${\bxi}$ is a mode's displacement vector corresponding to eigenfrequency $\omega$. The matrix elements $\mathcal{V}$, $\mathcal{T}$, $\mathcal{W}$ are operators defining the potential energy, kinetic energy, and Coriolis force. For two modes indexed by $\alpha$ and $\beta$, the corresponding matrix elements are 
\beq
\label{eq:tmat}
T_{\alpha \beta} = \int dV \rho \bxi_\alpha^* \cdot \bxi_\beta \, ,
\eeq
\begin{align}
\label{eq:wmat}
W_{\alpha \beta} &= \int dV \rho \bxi_\alpha^* \cdot \big( i {\boldsymbol \Omega} \times \bxi_\beta \big) \nonumber \\
&= -m \Omega \delta_{m_\alpha m_\beta} \int dr \rho r^2 \big( \xi_\perp^2 + 2 \xi_\perp \xi_r \big) \nonumber \\
&= - m \Omega \delta_{m_\alpha m_\beta} C_\alpha \, ,
\end{align}
where $\xi_{\perp}$ and $\xi_r$ are the horizontal and radial component of the mode's displacement vector, and the integral in equation \ref{eq:wmat} is equal to $C_\alpha$. The second line of equation \ref{eq:wmat} indicates modes only couple with the same value of $m$ due to the axisymmetric nature of the Coriolis force.

The expression for $V_{\alpha \beta}$ is given by equation 7.36 of \cite{dahlen:98}. These operators are real and symmetric such that $T_{\alpha \beta} = T_{\beta \alpha}$ and likewise for $V_{\alpha \beta}$ and $W_{\alpha \beta}$. Without rotation or tidal perturbations, equation \ref{eq:mat1} reduces to $\mathcal{V} \bxi_\alpha = \omega_\alpha^2 \mathcal{T} \bxi_\alpha$ for a mode of index $\alpha$ with eigenfrequency $\omega_\alpha$, with the orthonomality requirement $T_{\alpha \beta} = \delta_{\alpha \beta}$. In other words, we use the normalization convention $\int dr \rho r^2 \big( \xi_r^2 + \ell(\ell+1) \xi_\perp^2 \big) = 1$.

Our next step is to include the Coriolis force as a perturbation such that $W_{\alpha \beta} \ll \omega T_{\alpha \beta}$. We will solve for the new eigenfrequencies $\omega + \delta \omega$ for small perturbations $|\delta \omega| \ll \omega$. Then the bottom row of equation \ref{eq:mat1} yields
\beq
\delta \omega \simeq W_{\alpha \alpha} \, .
\eeq
This is the usual expression for the frequency perturbation induced by the Coriolis force. Since the Coriolis force only couples modes of the same $m$, modes of a rotating star are spherical harmonics of a unique value of $m$.

We next add the effects of tidal perturbations, which perturb the operators $\mathcal{T} \rightarrow \mathcal{T} + \delta \mathcal{T}$ and $\mathcal{V} \rightarrow \mathcal{V} + \delta \mathcal{V}$. Keeping only zeroth order and first order terms, the bottom row of equation \ref{eq:mat1} can be written 
\beq
\label{eq:pert}
\big( \omega_\alpha^2 \mathcal{T} + \delta \mathcal{V} - \omega_\alpha ^2 \delta \mathcal{T} + 2 \omega_\alpha \mathcal{W} \big) \bxi \simeq \omega^2 \mathcal{T} \bxi \, .
\eeq
We shall see that the operators $\delta \mathcal{T}$ and $\delta \mathcal{V}$ couple spherical harmonics of different $m$, so the eigenvectors of this system will no longer be individual spherical harmonics, but some superposition of spherical harmonics with different $m$.

To account for this coupling, we expand the eigenvectors $\bxi$ in terms of spherical harmonics of different $m$ such that
\beq
\bxi = \sum_m a_m \bxi_{\alpha,m} \, .
\eeq
Here, $a_m$ is the relative amplitude of each $m$ component to the eigenfunction, and our goal is to compute the values of $a_m$ for each eigenmode. We insert this expansion into equation \ref{eq:pert}, multiply each side by $\bxi^*$, and integrate over the volume of the star to obtain
\begin{align}
\label{eq:summ}
&\sum_m \big( \omega_\alpha^2 \delta_{m m'} + \delta V_{\alpha,m m'} - \omega_\alpha ^2 \delta T_{\alpha,m m'} + 2 \omega_\alpha W_{\alpha,m m'} \big) a_m  \nonumber \\ &\simeq \omega^2 \sum_m a_m \delta_{m m'} \, .
\end{align}
In the following sections, we will solve this equation for coupled dipole modes and quadrupole modes. Since we are only considering coupling within a mode multiplet, we drop the $\alpha$ subscripts on the matrix elements $\delta T_{\alpha,mm'}$, $\delta V_{\alpha,mm'}$, $\delta W_{\alpha,mm'}$. 

The values of the coupling coefficients in the above equations are computed following \cite{fuller:20} who builds on \cite{dahlen:98}. The kinetic energy perturbation due to aspherical distortion has form
\begin{equation}
\label{eq:dt}
    \delta T_{mm'} = \sum_{\rm S} X_{\ell 2 \ell'}^{m m_t m'} \int^{R}_0 \frac{2}{3} \varepsilon_{\rm S} \rho r^2 \Big[ \bar{T} - (\eta +3) \check{T} \Big] dr 
\end{equation}
where $\varepsilon_{\rm S}$ is the ellipticity, and $\eta$ is its radial derivative
\begin{align}
\eta &= \frac{\partial \ln \varepsilon}{\partial \ln r} = 3 - \frac{4 \uppi \rho r^3}{m(r)} \, .
\end{align}
We must sum over each source ${\rm S}$ of distortion, i.e., each component of the tidal potential and centrifugal distortion. We are only considering $\ell=2$ components of the tidal and centrifugal distortion, so this distortion is always quadrupolar in shape.  The $\bar{T}$ and $\check{T}$ terms depend on the mode eigenfunction $\bxi_\alpha$ and are  provided in \cite{fuller:20}.

The $X_{\ell \ell_t \ell}^{m m_t m'}$ coefficient is a geometric factor accounting for the angular overlap between the two coupled modes and quadrupolar asphericity. We are considering coupling induced by a distortion of angular pattern ($\ell_t$,$m_t$) between two modes of index $\ell$ and pattern numbers $m$ and $m'$. Its value is
\begin{align}
&X_{\ell \ell_t \ell}^{m m_t m'} = \bigg(\frac{4 \uppi}{2 \ell_t +1}\bigg)^{1/2} \int d \Omega \, Y_{\ell m}^* Y_{\ell_t m_t} Y_{\ell' m'} \nonumber \\
& = (-1)^m (2 \ell + 1) \begin{pmatrix}
\ell & \ell_t & \ell \\
-m & m_t & m'
\end{pmatrix}
\begin{pmatrix}
\ell & \ell_t & \ell\\
0 & 0 & 0
\end{pmatrix}
\end{align}
and the terms in large parenthesis are Wigner 3j symbols.

The potential energy coupling terms are 
\begin{align}
\label{eq:dV}
\delta V_{m m'} &= \sum_{\rm S} X_{\ell 2 \ell'}^{m m_t m'} \int^R_0 \frac{2}{3} \varepsilon r^2 dr \nonumber \\
&\times \bigg( \rho c_s^2 \big[\bar{V}_\kappa - (\eta +1) \check{V}_\kappa \big] + \rho \big[ \bar{V}_\rho - (\eta + 3) \check{V}_\rho \big] \bigg) 
\end{align}
where $c_s$ is the adiabatic sound speed. Again, the $V_\kappa$ and $V_\rho$ terms depend on $\bxi_{\alpha}$ and are provided in \cite{fuller:20}.

To account for all sources of asphericity, we begin with the tidal distortion, whose dominant components $U_{\ell_t m_t}$ from the $\ell_t=2$ component of the tidal potential $U$ are
\begin{align}
\label{eq:u}
    U_{\ell_t} &= \sum_{m_t} U_{\ell_t m_t} = U_{2-2} + U_{20} + U_{22} \nonumber \\
    &= \bigg( - \sqrt{\frac{3 \uppi}{10}} \Big[ Y_{2-2} + Y_{22} \Big] + \sqrt{\frac{\uppi}{5}} Y_{20}\bigg) \frac{G M_{\rm c} r^2}{a^3} \, .
\end{align}
Here, $M_c$ is the mass of the companion star, and $a$ is the orbital semi-major axis.

For each tidal component, we can compute the effective ellipticity $\varepsilon_{\rm S}$ as follows. The ellipticity is defined in \cite{dahlen:98} via the associated radial displacement
\begin{equation}
    \xi_{r,{\rm S}} = -\frac{2}{3} \varepsilon_{\rm S} \sqrt{\frac{4 \uppi}{2 \ell_t +1}} Y_{l_t m_t} r \, .
\end{equation}
For tidal or centrifugal distortion due to a potential $U_{l_t m_t}$, the radial displacement is $\xi_{r,{\rm S}} = - U_{l_t m_t}/g$, where $g=GM/r^2$, when the Cowling approximation is used to simplify the calculation. The effective ellipticity for the $l_t=2$, $m_t=0$ component is thus
\begin{equation}
\label{eq:epsval}
    \frac{2}{3} \varepsilon = \frac{1}{2} \frac{M_c}{M} \frac{r^3}{a^3} \, .
\end{equation}
For this component, we thus find
\begin{align}
\label{eq:dtex}
    \delta T_{mm'} &= \frac{1}{2} X_{\ell 2 \ell}^{m 0 m'} \frac{M_c}{M} \frac{R^3}{a^3} \int^{R}_0 \rho r^2 \Big[ \bar{T} - (\eta +3) \check{T} \Big]  \frac{r^3}{R^3}  dr \nonumber \\
    &= \frac{1}{2} X_{\ell 2 \ell}^{m 0 m'} \epsilon \ T_{\rm int}.
\end{align}
The second line defines the dimensionless tidal distortion 
\begin{equation}
\epsilon = \frac{M_c}{M} \bigg(\frac{R}{a} \bigg)^3
\end{equation}
with $T_{\rm int}$ equal to the integral in the first line. Similarly,
\begin{equation}
\label{eq:dvex}
    \delta V_{m m'} = \frac{1}{2} X_{\ell 2 \ell}^{m 0 m'} \epsilon \ V_{\rm int}
\end{equation}
where $V_{\rm int}$ is the integral in equation \ref{eq:dV}, with $(2/3) \varepsilon$ replaced by $(r/R)^3$. 

Note that the axisymmetric nature of the $m_t=0$ component of the tidal potential means that $X_{\ell 2 \ell}^{m 0 m'}$ is non-zero only for $m=m'$, i.e., it induces self-coupling that perturbs the mode frequencies, but it does not couple different values of $m$ together.

In contrast, the $m_t = \pm 2$ components of the tidal potential do induce such coupling. Following the same procedure, its form is 
\begin{equation}
\delta T_{mm'} = -\frac{1}{2} \sqrt{\frac{3}{2}} X_{\ell 2 \ell}^{m \pm2 m'} \epsilon \ T_{\rm int} \, ,
\end{equation}
and similarly for $\delta V_{mm'}$. The factor of $-\sqrt{3/2}$ comes from the different coefficient in front of the $m_t = \pm2$ components of the tidal potential in equation \ref{eq:u}. This couples $m=1$ modes with $m=-1$ modes, producing off-diagonal terms in the matrices discussed below that are crucial for changing the mode geometry.

The centrifugal distortion due to the star's rotation also introduces an $\ell_t=2$, $m_t=0$ distortion that adds to that produced by the tidal force. We can determine its form by writing the centrifugal potential
\begin{align}
    U^{\rm cen} &= - \frac{1}{2} \Omega^2 R^2 \nonumber \\
    &= \bigg[ \frac{1}{3} \sqrt{\frac{4 \uppi}{5}} Y_{20} - \frac{1}{3} \sqrt{4 \uppi} Y_{00} \bigg] \Omega^2 r^2 \, .
\end{align}
Since we assume the pulsating star is tidally synchronized such that $\Omega^2 = G(M+M_c)/a^3$, we have
\begin{equation}
\label{eq:ucen}
    U^{\rm cen}_{\ell_t=2} = \frac{2}{3} \frac{M+M_c}{M_c} U^{\rm tide}_{\ell_t=2,m_t=0} \, .
\end{equation}
Hence, the centrifugal form of $\delta T$ and $\delta V$ is the same as that of the $\ell_t=2$, $m_t=0$ component of the tidal potential, but with an amplitude different by a factor of $(2/3)(M+M_c)/M_c$. 

Finally, the $\ell_t=m_t=0$ component of the centrifugal distortion also induces self-coupling that perturbs the mode eigenfrequencies. Following \cite{dahlen:98}, the perturbation element is
\begin{equation}
\label{eq:dvcen}
    \delta V_{\rm cen} = \frac{2\Omega^2}{3} \bigg[ 1 - \ell(\ell+1) \int^R_0 \rho r^2  \big(\xi_\perp^2 + 2 \xi_{\perp} \xi_r \big) dr \bigg] \, .
\end{equation}

In the following sections, we combine each of these effects to provide the explicit form of the matrix equations that must be solved to obtain the new mode frequencies and eigenfunctions in the presence of tidal distortion.

\subsection{Dipole Modes}
\label{sec:dipole}

For dipole modes, the coupled mode eigensystem of equation \ref{eq:summ} can be written in matrix form
\begin{align}
\label{eq:matrixcoup}
&
\begin{bmatrix}
\omega_\alpha^2 + \delta \omega^2_{-1-1} & 0 & \delta \omega^2_{-11} \\
0 & \omega_\alpha^2 + \delta \omega^2_{00} & 0 \\
\delta \omega^2_{1 -1} & 0 & \omega_\alpha^2 + \delta \omega^2_{11}
\end{bmatrix}
\begin{bmatrix}
a_{-1} \\
a_0  \\
a_{1} 
\end{bmatrix}
\newline \nonumber \\
&= \omega^2
\begin{bmatrix}
a_{-1} \\
a_0  \\
a_{1} 
\end{bmatrix}
\end{align}
where the $\delta \omega_{m m'}^2$ are perturbations produced by the combination of tidal, centrifugal, and the Coriolis force as described above. The $a_m$ values are the relative contributions of the $Y_{1m}$ components to the new tidally modified eigenfunctions.

The values of $\delta \omega_{mm'}$ are the perturbative terms in equation \ref{eq:summ}, with $\delta T_{mm'}$ from equation \ref{eq:dtex}, $\delta V_{mm'}$ from equation \ref{eq:dvex}, and $W_{\rm int}$ is the integral in equation \ref{eq:wmat}. Explicitly, we find
\begin{align}
    \delta \omega_{-1-1}^2 &= -\frac{1}{10} \epsilon \bigg(1 + \frac{2}{3}\frac{M+M_c}{M_c} \bigg) \big( V_{\rm int} - \omega_\alpha^2 \ T_{\rm int} \big) \nonumber \\ &+ \delta V_{\rm cen} + 2 \Omega \omega_\alpha W_{\rm int} \, .
\end{align}
The factor $1+2(M+M_c)/3M_c$ arises from adding both the tidal and centrifugal distortion (via equation \ref{eq:ucen}), and the factor $\delta V_{\rm cen}$ of equation \ref{eq:dvcen} comes from the spherical component of the centrifugal force. The expression for $\delta \omega_{11}^2$ is identical, apart from the sign of the last Coriolis term. We used the fact that $X_{1 2 1}^{1 0 1} = -1/5$.  Similarly, using $X_{1 2 1}^{0 0 0} = 2/5$, we have
\begin{align}
    \delta \omega_{00}^2 &= \frac{1}{5} \epsilon \bigg(1 + \frac{2}{3}\frac{M+M_c}{M_c} \bigg) \big( V_{\rm int} - \omega_\alpha^2 \ T_{\rm int} \big) \nonumber \\
    &+ \delta V_{\rm cen} 
\end{align}
And finally, using $-\sqrt{\frac{3}{2}} X_{1 2 1}^{1 2 -1} = 3/5$, we have
\begin{align}
    \delta \omega_{1-1}^2 = \delta \omega_{-11}^2 &= \frac{3}{10} \epsilon \big( V_{\rm int} - \omega_\alpha^2 \ T_{\rm int} \big)  \, .
\end{align}

Solving the eigensystem of equation \ref{eq:matrixcoup} yields the three mode frequencies $\omega^2$ and eigenfunctions ${\bf a}$ of the perturbed mode triplet. Since the $m=0$ mode remains uncoupled to the $m=\pm1$ components, its eigenfunction remains unchanged. It is symmetric about the $z-$axis, i.e., the spin/orbital axis, and we will refer to this as the $Y_{10z}$ mode. The perturbed eigenfrequency is 
\begin{equation}
    \omega_0 = \sqrt{ \omega_\alpha^2 + \delta \omega_{00}^2} \simeq \omega_\alpha + \frac{\delta \omega_{00}^2}{2 \omega_\alpha} \, .
\end{equation}

The $m=\pm1$ modes are coupled by the tidal distortion, forming a 2$\times$2 matrix equation, which can be easily solved. In the limit of weak tidal coupling such that the off diagonal terms are small, i.e., $\delta \omega_{\rm 1-1}^2 \ll|\delta \omega_{11}^2 - \delta \omega_{-1-1}^2| = |4 \Omega \omega_\alpha C_\alpha|$, we simply obtain modes with perturbed frequencies and eigenvectors ${\bf a} = [1, 0]$ and ${\bf a} = [0, 1]$ corresponding to the usual $m=1$ and $m=-1$ spherical harmonics. 

\begin{figure}
\includegraphics[scale=0.14]{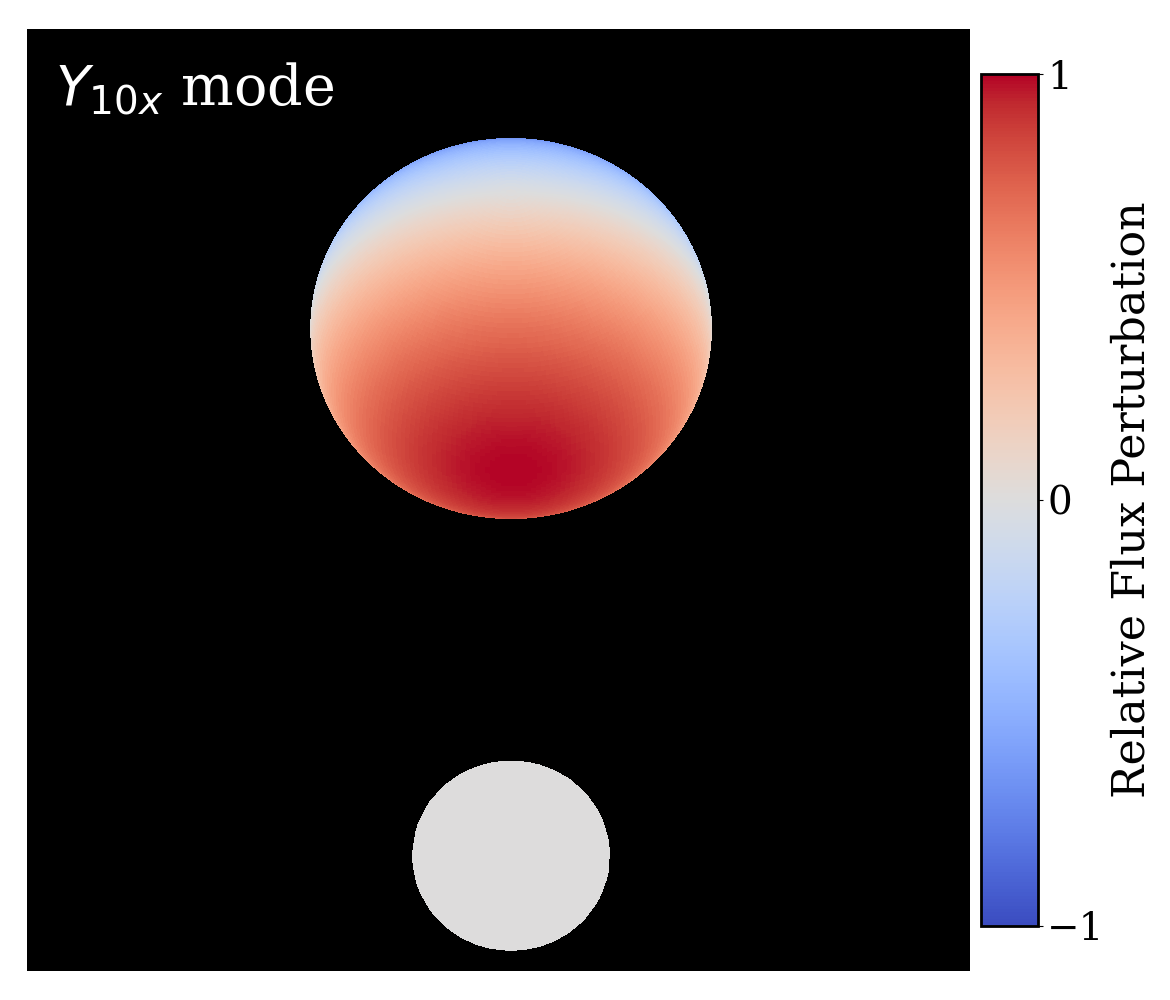}
\includegraphics[scale=0.14]{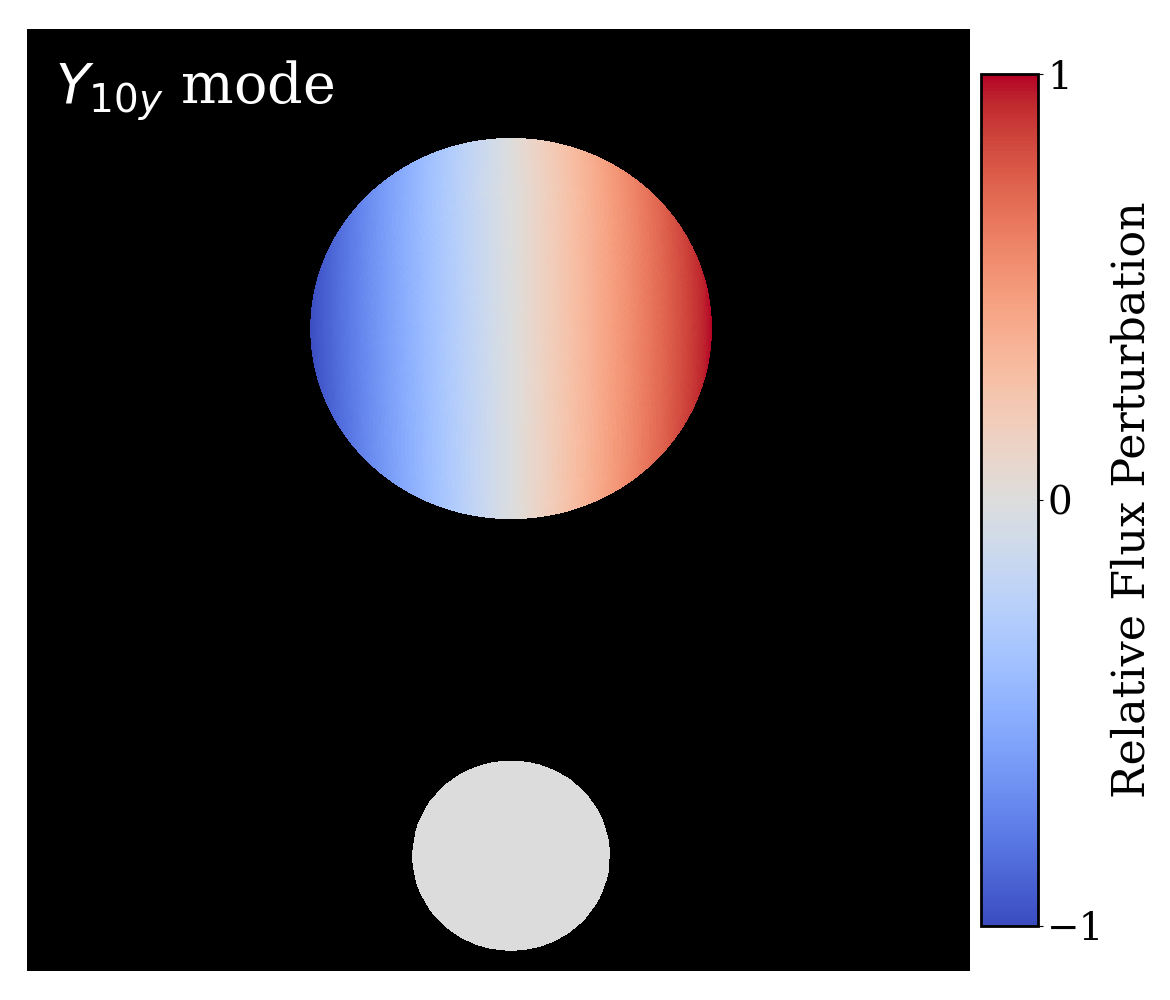}
\includegraphics[scale=0.14]{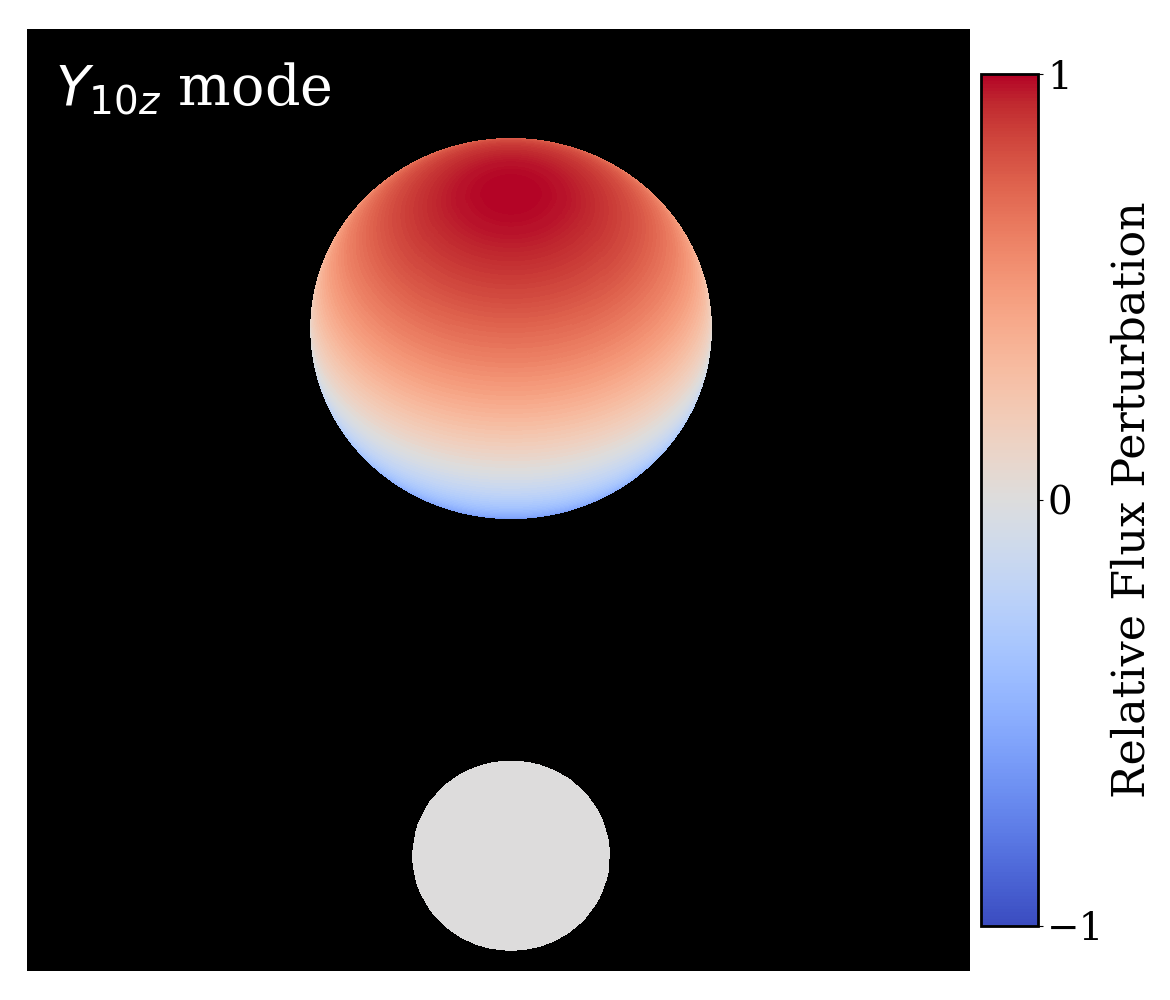}
\caption{\label{fig:dipole} Angular pattern of $\ell=1$ triaxial pulsations in a tidally distorted star, viewed at orbital inclination $i=45^\circ$ with the companion shown as a gray circle. The color indicates the relative amplitude of the surface flux perturbation (i.e., temperature perturbation) for the three $\ell=1$ pulsations: the $Y_{10x}$ mode aligned with the tidal axis (top left), the $Y_{10z}$ mode aligned with the spin axis (bottom left), and the $Y_{10y}$ mode perpendicular to both of these (top right). All modes are standing modes, so these patterns oscillate back and forth rather than propagating around the star. Movies showing these pulsations can be found \href{https://drive.google.com/drive/folders/1IwqleGn2IJPTbLWX8mddoLwL-2Hz24pH?usp=sharing}{here}.
}
\end{figure}

In the limit of strong tidal coupling such that $\delta \omega_{\rm 1-1}^2 \gg |\delta \omega_{11}^2- \delta \omega_{-1-1}^2|$, the new mode eigenfrequencies become 
\begin{equation}
    \omega_{\pm}^2 = \omega_\alpha^2 + \frac{1}{2} \bigg(\delta \omega_{-1-1}^2 + \delta \omega_{11}^2 \bigg) \pm \delta \omega_{\rm 1-1}^2 \, ,
\end{equation}
with corresponding eigenvectors 
\begin{equation}
{\bf a}_\pm =
\begin{bmatrix}
1 \\
\pm 1 
\end{bmatrix} \, .
\end{equation}
These frequencies are slightly different from those in \cite{zhang:24} and \cite{jayaraman:24}, since we have kept tidal perturbation terms from the diagonal elements.

Hence, the new mode eigenfunctions are equal superpositions of $Y_{1,1}$ and $Y_{1,-1}$, with angular flux perturbation patterns
\begin{equation}
    \delta F_\pm \propto \big[ Y_{11}(\theta,\phi) \pm Y_{1-1}(\theta,\phi) \big] e^{- i \omega_\pm t} \, .
\end{equation}
Some algebra shows that the spatial/time dependence of these two modes are
\begin{align}
\label{eq:xi+}
    \delta F_+ &\propto \sin \theta \sin \phi \sin (\omega_+ t) \nonumber \\
    &\propto y \sin (\omega_+ t)
\end{align}
and
\begin{align}
\label{eq:xi-}
    \delta F_- &\propto \sin \theta \cos \phi \cos (\omega_- t) \nonumber \\
    &\propto x \cos (\omega_- t) \, .
\end{align}
Here, $x$ and $y$ are Cartesian coordinates relative to the center of the star, and we have defined the $x$-axis to lie along the tidal axis. These results correct a mistake in \cite{zhang:24}, in which the $\omega_-$ and $\omega_+$ solutions are switched with each other. Note also that the $m=0$ mode has
\begin{align}
\label{eq:xi0}
    \delta F_0 & \propto Y_{10}(\theta,\phi) \cos(\omega_0 t) \nonumber \\
    &\propto \cos \theta \cos (\omega_0 t) \nonumber \\
    &\propto z \cos (\omega_0 t) \, .
\end{align}

\begin{figure}
\includegraphics[scale=0.37]{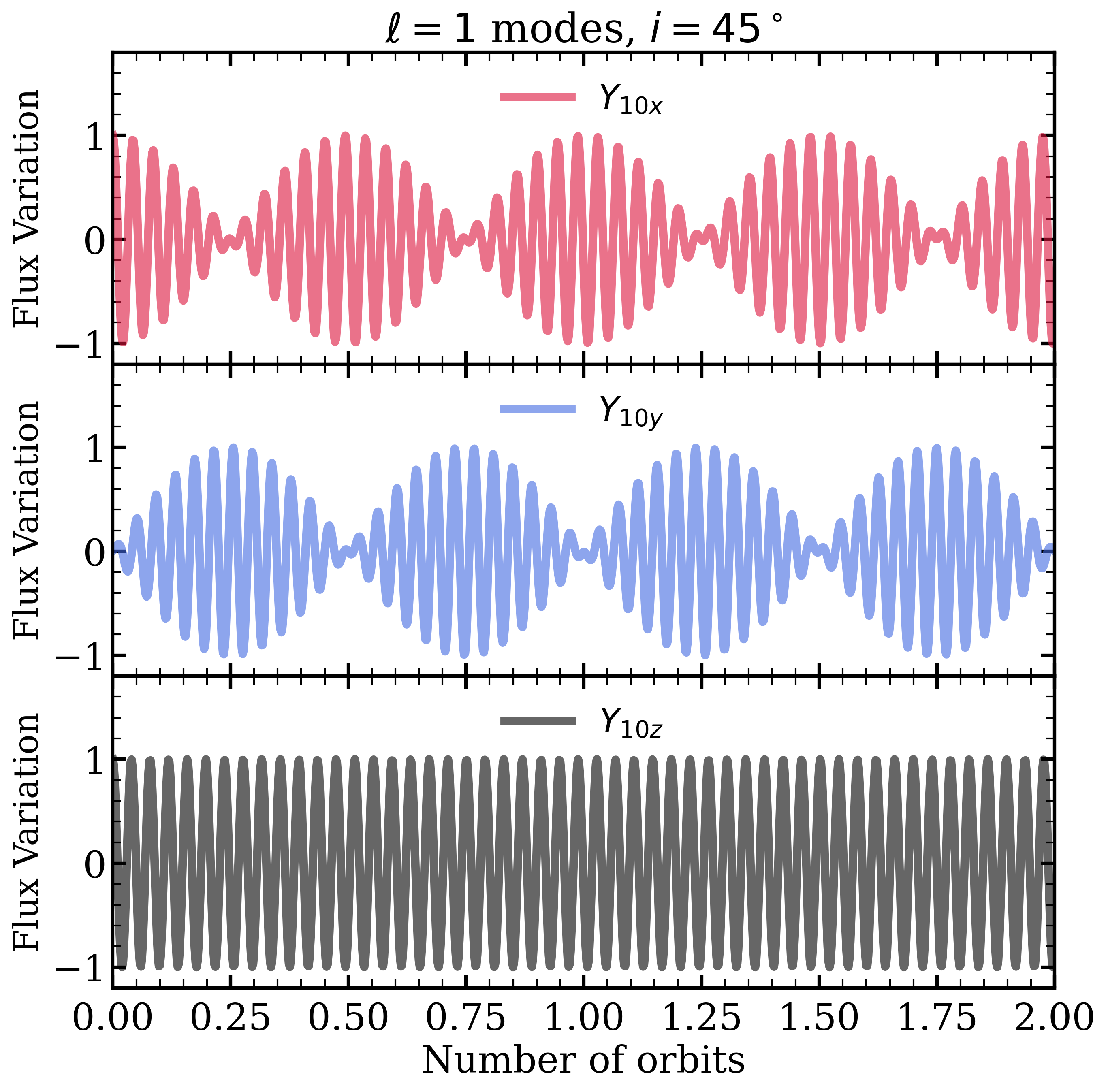}
\caption{\label{fig:l1lightcurve} Observed brightness variation produced by tidally tilted dipole modes, when viewed at orbital inclination $i=45^\circ$, showing the $Y_{10x}$ mode (top panel), $Y_{10y}$ mode (middle panel), and $Y_{10z}$ mode (bottom panel). This plot shows the $l=1$ p modes of radial order $n_{\rm pg} = 4$ for the model $\delta$\,Scuti pulsator in Section \ref{sec:application}. The amplitudes of $Y_{10x}$ and $Y_{10y}$ modes are modulated twice per orbit, but $Y_{10z}$ modes have constant amplitude. 
}
\end{figure}

Equations \ref{eq:xi+} and \ref{eq:xi-} show that strongly coupled $m=\pm1$ modes will create two new eigenmodes, both with largest displacements in the orbital plane. The $+$ mode has an eigenfunction peaked at $\phi = \pm \uppi/2$. It has the same eigenfunction as a $Y_{10}$ spherical harmonic, except aligned with the $y$-axis instead of the $z$-axis, so we refer to it as the $Y_{10y}$ mode. The $-$ mode has an eigenfunction peaked at $\phi=0$ and $\phi = \uppi$, i.e., it is aligned with the $x-$axis, so we refer to it as the $Y_{10x}$ mode. Neither mode propagates around the equator like uncoupled $m=\pm 1$ modes. Instead, both modes are standing modes, aligned with the $x-$ and $y-$axes. The $m=0$ mode is a standing mode aligned with the $z-$axis. Hence, tidally coupled $\ell=1$ modes are triaxial because each is aligned with one of the three major axes of star, which is a triaxial ellipsoid due to its tidal and centrifugal distortion. 

\begin{figure*}
\includegraphics[scale=0.62]{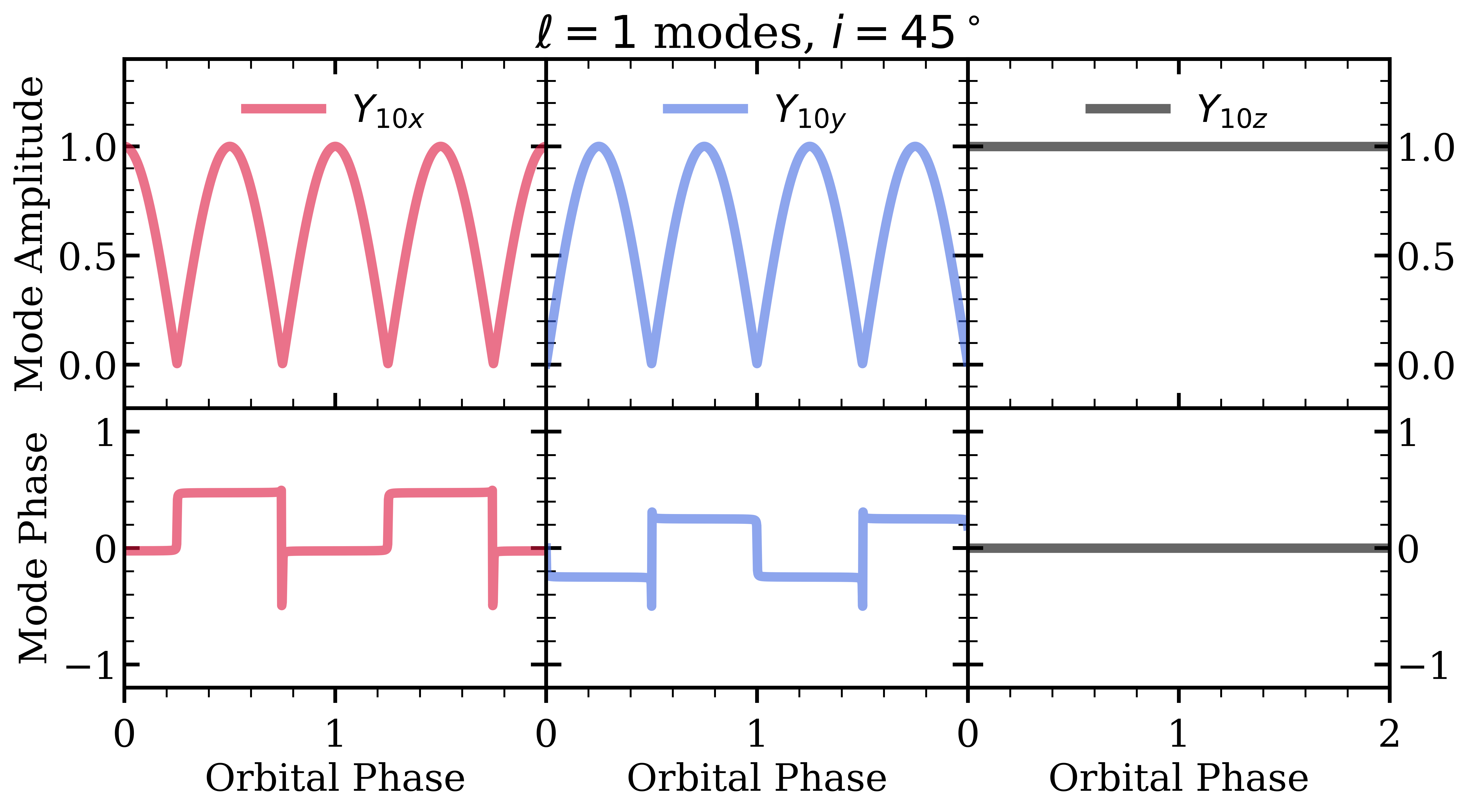}
\caption{\label{fig:l1amp} Amplitude (top panel) and phase variation (bottom panel) of tidally tilted dipole modes over the course of one orbit, for the same modes shown in Figure \ref{fig:l1lightcurve}. The $Y_{10x}$ modes peak at orbital phases $0$ and $0.5$, while the $Y_{10y}$ modes peak at orbital phases $0.25$ and $0.75$. These modes change phase by 0.5 cycles when their amplitude crosses zero.}
\end{figure*}

\subsubsection{Dipole mode visibility and amplitude modulation}

The geometry of triaxial pulsations determines the amplitude and phase modulation of the modes over the course of the orbit. Because the mode pattern is fixed relative to the tidal axis, it can vary relative to an external observer.

\begin{figure}
\includegraphics[scale=0.37]{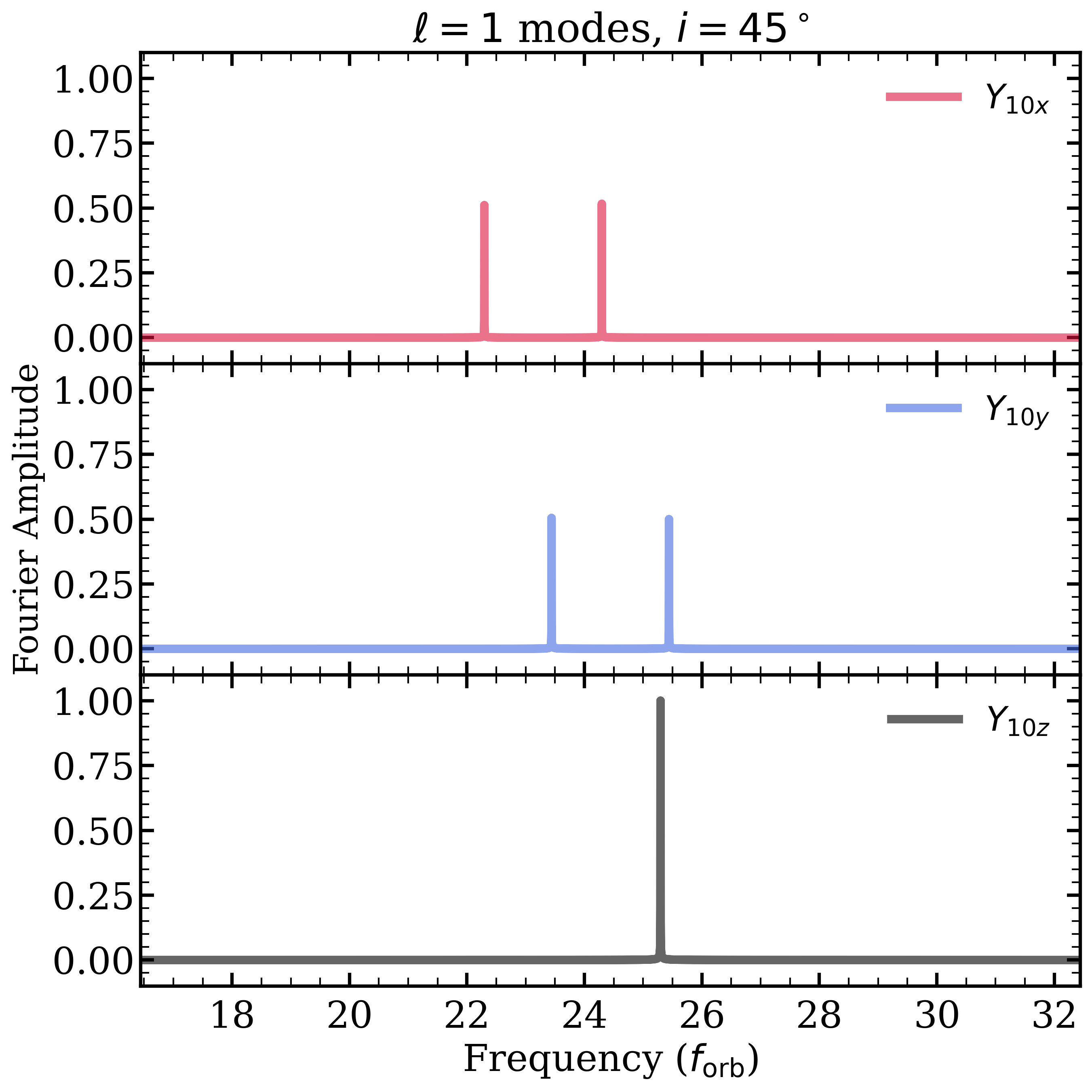}
\caption{\label{fig:l1fourier} Power spectrum of the tidally tilted dipole modes from Figures \ref{fig:l1lightcurve} and \ref{fig:l1amp}. The $Y_{10x}$ and $Y_{10y}$ modes produce equal-amplitude doublets spaced by exactly 2$\times$ the orbital frequency, while the $Y_{10z}$ mode produces a singlet. 
}
\end{figure}

We compute the pulsation light curves in the same manner as \cite{fuller:20}. For each mode, one could also simply sum up the contribution of the different $m$ components. For dipole modes, we show the resulting light curves (Figure \ref{fig:l1lightcurve}), amplitude and phase modulation (Figure \ref{fig:l1amp}), and observed amplitude spectrum (Figure \ref{fig:l1fourier}).  These plots show numerical solutions to equation \ref{eq:matrixcoup} without making the approximations discussed above. They are made using the $n_{\rm pg}=4$ mode of our $\delta$\,Scuti model from Section \ref{sec:application}, at an orbital period of $P_{\rm orb}=1\,{\rm day}$ and companion mass of $1.3\,{\rm M}_\odot$.

The $Y_{10x}$ mode is aligned with the tidal axis and produces the largest flux modulations when the tidal axis is closest to the line of sight, i.e., orbital phases 0 and 0.5. The $Y_{10y}$ mode is very similar, but shifted in orbital phase, such that its maximal amplitudes are at orbital phases 0.25 and 0.75. The $Y_{10z}$ mode exhibits no modulation because the $z$-axis (i.e., the orbital/spin axis) does not vary its angle with respect to the line of sight.

The amplitude of the $Y_{10x}$ and $Y_{10y}$ modes peaks twice per orbit, and the pulsation phase jumps by 0.5 cycles each time the amplitude crosses zero. In an amplitude spectrum, both modes produce two peaks, each with nearly equal amplitude, and spaced by exactly twice the orbital frequency. In contrast, the $Y_{10z}$ mode only produces a single peak in the power spectrum. Consequently, a dipole mode triplet produces five total peaks in a power spectrum, in contrast to the three peaks produced by modes aligned with the spin axis. Of course, not all modes may be excited to detectable amplitudes, complicating mode identification efforts. Additionally, the tidal frequency splitting can be greater than the large frequency spacing (see Figure \ref{fig:tidalsplitperechstar}), making it more difficult to confidently identify $\ell=1$ multiplets.

It is important to note that triaxial $\ell=1$ modes produce different amplitude and phase variability than $\ell=1$ modes aligned with the tidal axis. The triaxial modes we predict are always standing modes. In contrast, $\ell=1$ and $m=\pm1$ modes aligned with the tidal axis would be traveling modes around the tidal axis. This typically produces triplets in a power spectrum (see Figure 10 of \citealt{jayaraman:24}), which are not observed for most of the modes in that work or in \cite{zhang:24}.

\begin{figure}
\includegraphics[scale=0.14]{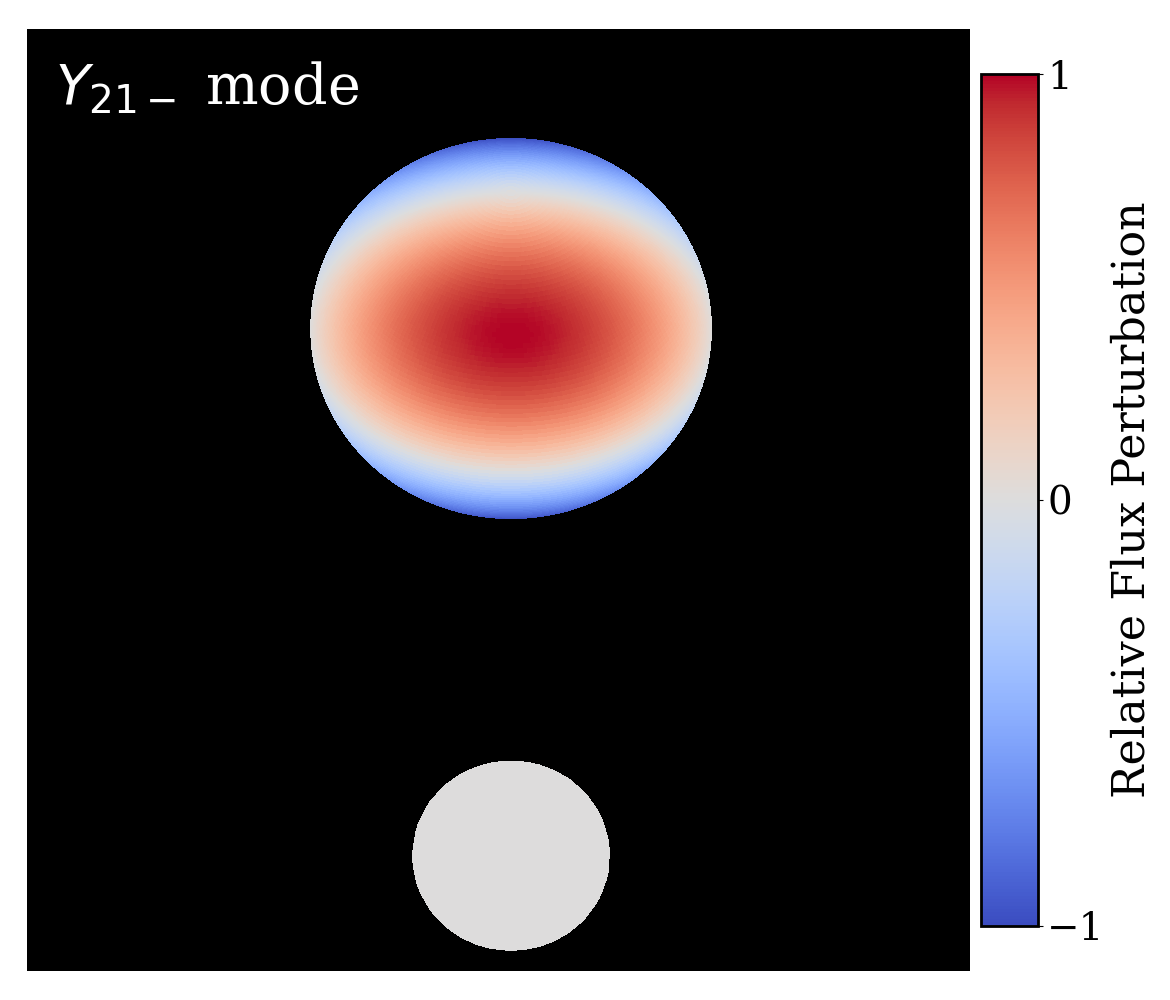}
\includegraphics[scale=0.14]{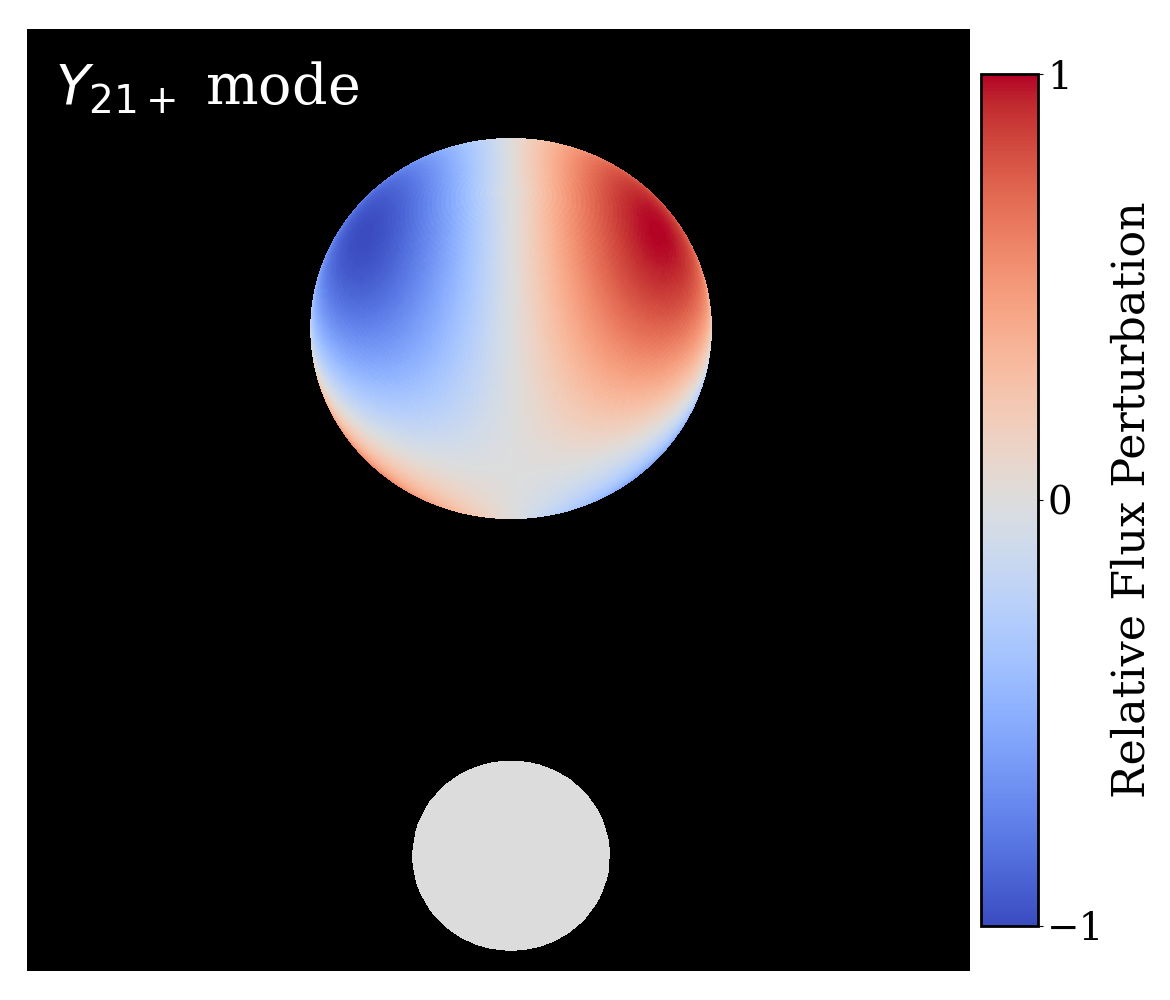}
\includegraphics[scale=0.14]{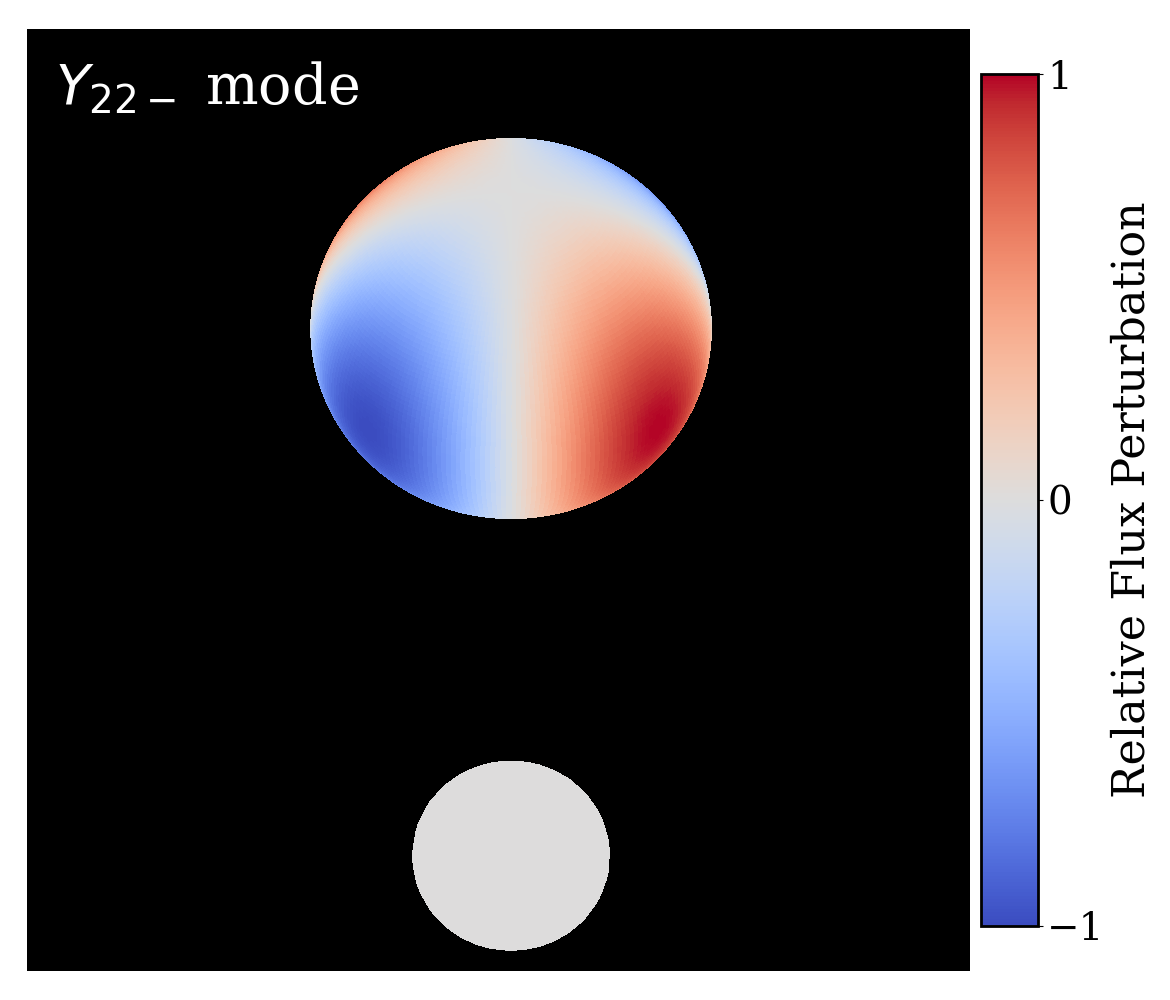}
\includegraphics[scale=0.14]{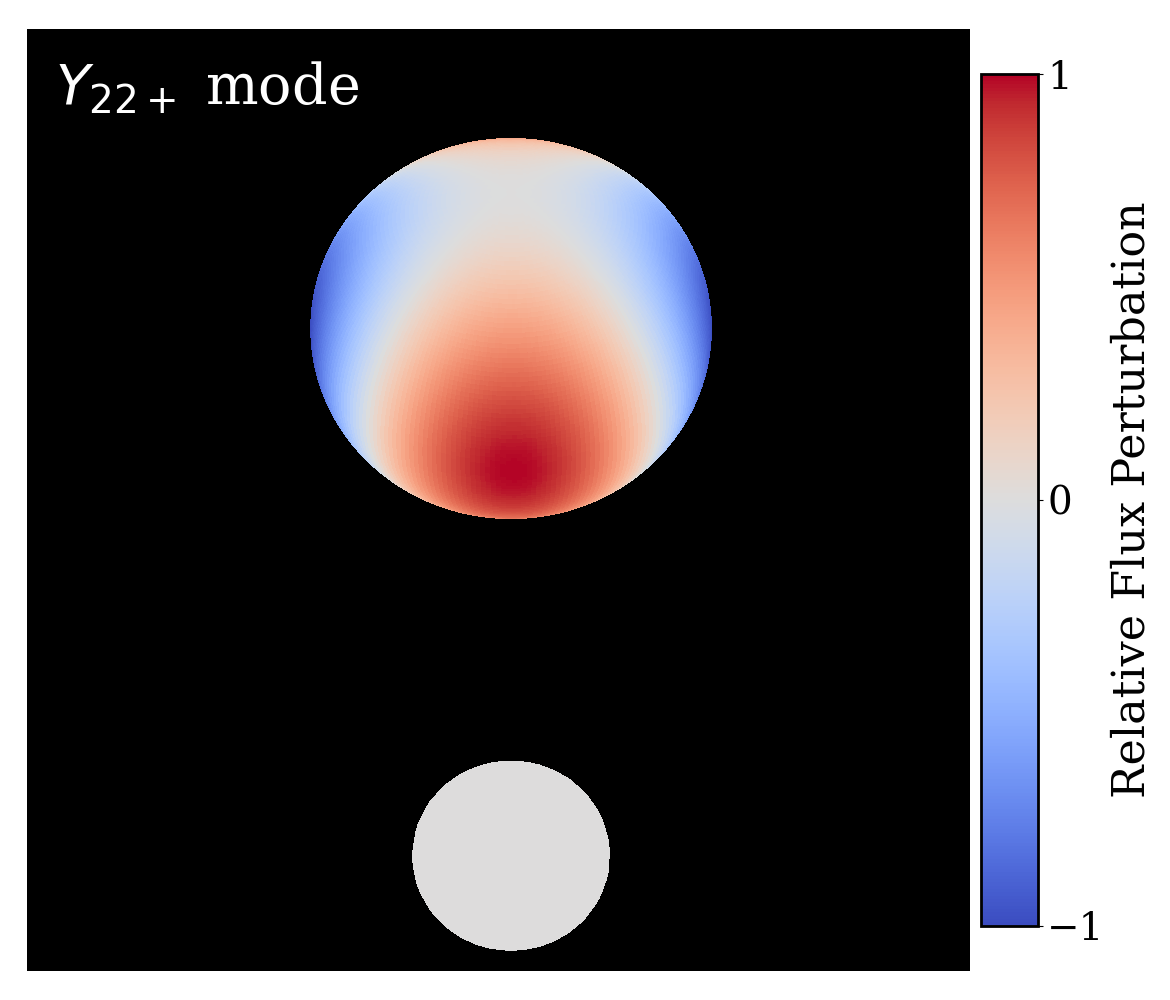}
\includegraphics[scale=0.14]{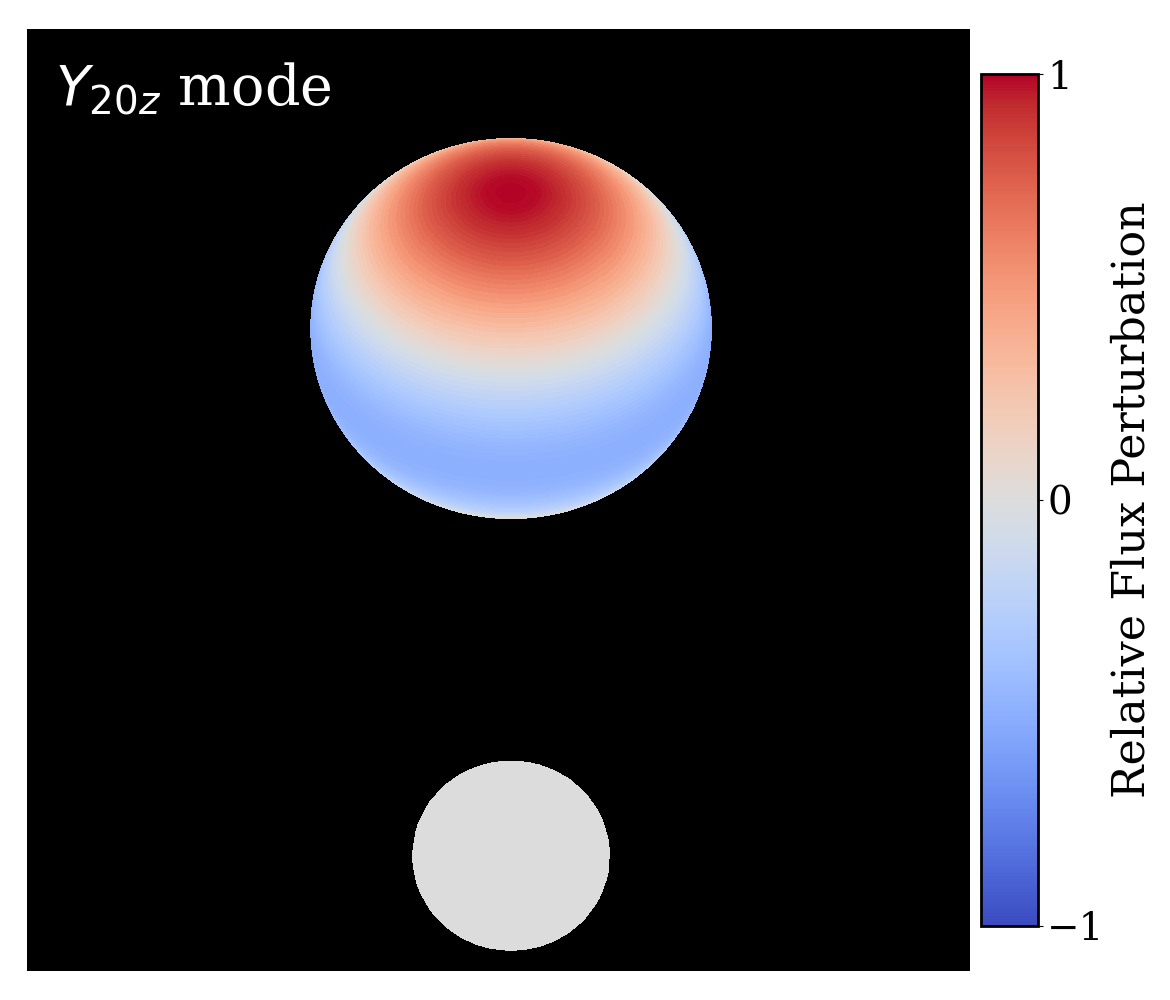}
\caption{\label{fig:quadrupole} Same as Figure \ref{fig:dipole}, but now for the five $\ell=2$ pulsations of a triaxial pulsator. These are the $Y_{21-}$ mode (top left), $Y_{21+}$ mode (top right), $Y_{22-}$ mode (middle left), the $Y_{22+}$ mode (middle right) and the $Y_{20z}$ mode (bottom left).
Once again, all modes are standing modes. Movies showing these pulsations can be found \href{https://drive.google.com/drive/folders/1IwqleGn2IJPTbLWX8mddoLwL-2Hz24pH?usp=sharing}{here}.
}
\end{figure}

\subsubsection{Accounting for the Coriolis force}

Above, we presented results in the limit of strong tidal coupling such that $\delta \omega_{\rm 1-1}^2 \ll|\delta \omega_{11}^2 - \delta \omega_{-1-1}^2| = |4 \Omega \omega_\alpha C_\alpha|$. If we repeat the same exercise but keep terms of first order in the Coriolis force, we obtain eigenfrequencies of
\begin{equation}
    \omega_{\pm}^2 \simeq \omega_\alpha^2 + \frac{1}{2} \bigg(\delta \omega_{-1-1}^2 + \delta \omega_{11}^2 \bigg) \pm \bigg[\delta \omega_{\rm 1-1}^2 + \frac{2 \Omega^2 \omega_\alpha^2 C_\alpha^2}{\delta \omega_{1-1}^2} \bigg] \, 
\end{equation}
and corresponding eigenvectors
\begin{equation}
{\bf a}_\pm \simeq
\begin{bmatrix}
1 \pm \frac{2 \Omega \omega_\alpha C_\alpha}{\delta \omega_{1-1}^2}  \\
\pm 1 
\end{bmatrix} \, .
\end{equation}

The corresponding flux variations have spatial and time dependence
\begin{equation}
    \delta F_\pm \propto \bigg[ Y_{11}(\theta,\phi) \pm \bigg(1 \pm \frac{2 \Omega \omega_\alpha C_\alpha}{\delta \omega_{1-1}^2} \bigg) Y_{1-1}(\theta,\phi) \bigg] e^{- i \omega_\pm t} \, .
\end{equation}
This can be rearranged to obtain
\begin{align}
\label{eq:xi+c}
    \delta F_+ &\propto y \sin (\omega_+ t) - \frac{2 \Omega \omega_\alpha C_\alpha}{\delta \omega_{1-1}^2} \cos (\phi - \omega_+ t) 
\end{align}
and
\begin{align}
\label{eq:xi-c}
    \delta F_- &\propto x \cos (\omega_- t) - \frac{2 \Omega \omega_\alpha C_\alpha}{\delta \omega_{1-1}^2} \cos (\phi + \omega_- t) \, .
\end{align}

Hence, the effect of sub-dominant the Coriolis force is that both the $Y_{10x}$ and $Y_{10y}$ modes obtain a small circulating component. In a power spectrum, each mode will still produce a doublet, but the amplitude of the peaks will be slightly different, with a ratio of $1 - 2 \Omega \omega_\alpha C_\alpha/\delta \omega_{1-1}^2$.

\subsection{Quadrupole modes}
\label{sec:quad}

We next apply our calculation to the five quadrupole oscillation modes of a multiplet. Since the tidal force only has $m=0$ and $m=2$ components, the even $m$ and odd $m$ modes (i.e., those symmetric and anti-symmetric across the equator) will be uncoupled to one another. 

The $\ell=2$, $m=-1$ and $m=1$ modes will be coupled in a very similar fashion to the $\ell=1$ modes discussed above. They form a matrix equation very similar to equation \ref{eq:matrixcoup}, except that the values of the matrix elements are slightly different due to different Wigner-3j symbols arising from the angular overlap integrals. Once again, in the limit that $\delta \omega_{\rm 1-1}^2 \gg |\delta \omega_{-1-1}^2-\delta\omega_{11}^2|$, we have 
\begin{equation}
    \omega_{\pm}^2 = \omega_\alpha^2 \pm \delta \omega_{\rm 1-1}^2 \, ,
\end{equation}
with corresponding eigenvectors ${\bf a} = [0, 1, 0, \pm 1, 0]$ where the vector ${\bf a}$ now gives the amplitudes of the components $[Y_{2-2}$,$Y_{2-1}$,$Y_{20}$,$Y_{21}$,$Y_{22}]$. 

The corresponding eigenfunctions are equal superpositions of $Y_{21}$ and $Y_{2-1}$, with angular flux perturbation patterns
\begin{equation}
    \delta F_\pm \propto \big[ Y_{21}(\theta,\phi) \pm Y_{2-1}(\theta,\phi) \big] e^{- i \omega_\pm t} \, .
\end{equation}
Some algebra shows that the spatial/time dependence of these two modes are
\begin{align}
\label{eq:xi+2}
    \delta F_+ &\propto \sin \theta \cos \theta \sin \phi \cos (\omega_+ t) \nonumber \\
    &\propto y z \cos (\omega_+ t)
\end{align}
and
\begin{align}
\label{eq:xi-2}
    \delta F_- &\propto \sin \theta \cos \theta \cos \phi \sin (\omega_- t) \nonumber \\
    &\propto x z \cos (\omega_- t) \, .
\end{align}
Both of these modes are standing modes, similar to the $\ell=1$ modes. These modes have a similar pattern to normal $\ell=2$, $m=1$ modes, (see Figure \ref{fig:quadrupole}), except that they are standing modes that do not propagate around the $z$-axis. We refer to them as the $Y_{21-}$ and $Y_{21+}$ modes, respectively. The only difference between the two eigenfunctions is their phase: the former has maxima/minima along the $x$-axis, while the latter has maxima/minima along the $y$-axis.

Next we examine the $m=2$, $m=0$, and $m=-2$ modes, which form a coupled triplet. The eigensystem for these modes is
\begin{equation}    
\label{eq:matrixcoupl2}
\begin{bmatrix}
\omega_\alpha^2 + \delta \omega^2_{-2-2} & \delta \omega_{\rm -2 0}^2 & 0 \\
\delta \omega_{\rm 0 -2}^2 & \omega_\alpha^2 + \delta \omega^2_{00} & \delta \omega_{\rm 0 2}^2 \\
0 & \delta \omega_{\rm 2 0}^2 & \omega_\alpha^2 + \delta \omega^2_{22} \\
\end{bmatrix}
\begin{bmatrix}
a_{-2} \\
a_0 \\
a_{2} \\
\end{bmatrix}
\simeq \omega^2
\begin{bmatrix}
a_{-2} \\
a_0 \\
a_{2} \\
\end{bmatrix} \, .
\end{equation}

Note that all the off-diagonal tidal coupling terms are equal to each other. 
However, in this case we cannot assume that $\delta \omega_{\rm 20}^2 \gg |\delta \omega_{22}^2-\delta\omega_{00}^2|$ since all of these terms are comparable to one another. Explicit evaluation of these terms gives
\beq
\delta \omega^2_{00} \simeq \frac{1}{7} \epsilon \bigg(1 + \frac{2}{3} \frac{M+M_c}{M_c}\bigg) \big(V_{\rm int} - \omega_\alpha^2 T_{\rm int}\big)
\eeq
\beq
\delta \omega^2_{22} \simeq \delta \omega^2_{2-2} \simeq -\frac{1}{7} \epsilon \bigg(1 + \frac{2}{3} \frac{M+M_c}{M_c}\bigg) \big(V_{\rm int} - \omega_\alpha^2 T_{\rm int}\big)
\eeq
\beq
\delta \omega^2_{20} \simeq \frac{1}{7} \sqrt{\frac{3}{2}} \epsilon  \big(V_{\rm int} - \omega_\alpha^2 T_{\rm int}\big) \, ,
\eeq
where we have neglected the Coriolis and centrifugal terms because they are small for p~modes. We thus see that $(\delta \omega_{00}^2-\delta\omega_{22}^2)/\delta \omega_{\rm 20}^2 \simeq 2 \sqrt{2/3} \big[1+2(M+M_c)/3M_c \big]$, which is always larger than one, and can be very large when $M_c \ll M$. 

Consequently, we focus on the case where $(\delta \omega_{00}^2-\delta\omega_{22}^2)/\delta \omega_{\rm 20}^2 \gg 1$, i.e., the off-diagonal terms are small relative to the difference between the diagonal terms, but we still assume the Coriolis perturbation terms are small relative to the tidal ones. Solving the matrix equation of \ref{eq:matrixcoupl2} in this limit yields three eigenvalues and eigenvectors. 
The first solution has 
\begin{equation}
    \omega^2 \simeq \omega_\alpha^2 + \delta \omega_{22}^2
\end{equation}
and eigenvector ${\bf a} = [1, 0, 0, 0, -1]$. The spatial dependence is
\begin{equation}
    \delta F \propto \big[-Y_{22}(\theta,\phi) + Y_{2-2}(\theta,\phi) \big] e^{- i \omega_0 t} \, .
\end{equation}
Some algebra shows this reduces to
\begin{align}
\label{eq:dfa}
    \delta F &\propto \sin^2 \theta \sin \phi \cos \phi \sin (\omega_0 t) \nonumber \\
    &\propto x y \sin (\omega_0 t) \, .
\end{align}
This mode has an $\ell=m=2$ pattern that is wrapped around the $z$-axis, with maxima/minima offset from the tidal axis by an angle of $\uppi/4$, and it is a standing mode that does not propagate around the $z$-axis. We refer to it as the $Y_{22-}$ mode.

\begin{figure}
\includegraphics[scale=0.37]{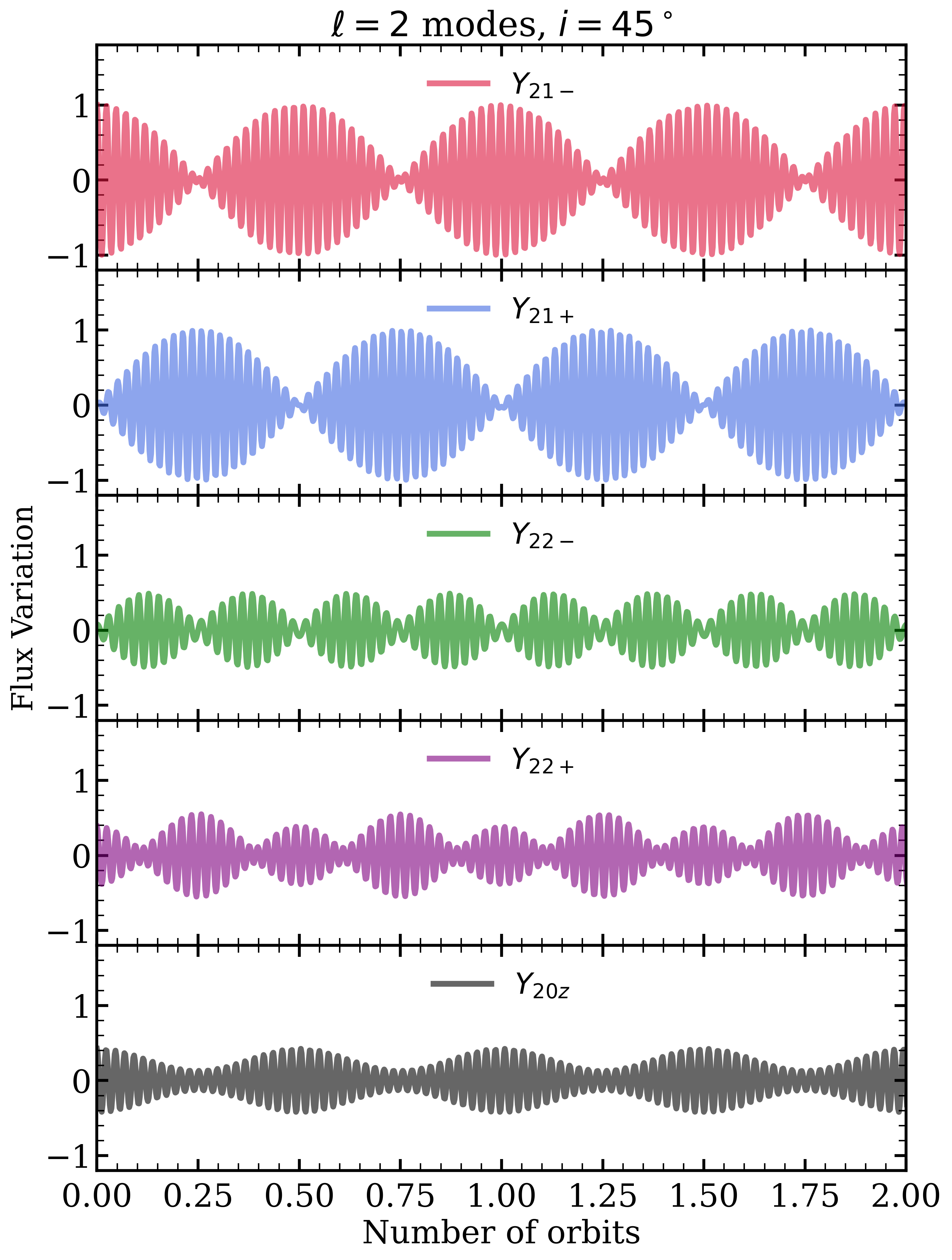}
\caption{\label{fig:l2lightcurve} Same as Figure \ref{fig:l1lightcurve}, but now for the five different triaxial $\ell=2$ pulsation modes. These modes correspond to radial order $n_{\rm pg}=8$ for the model of Section \ref{sec:application}. For $\ell=2$ modes, departures from the idealized triaxial modes computed in Section \ref{sec:quad} are more apparent.}
\end{figure}

\begin{figure*}
\includegraphics[scale=0.62]{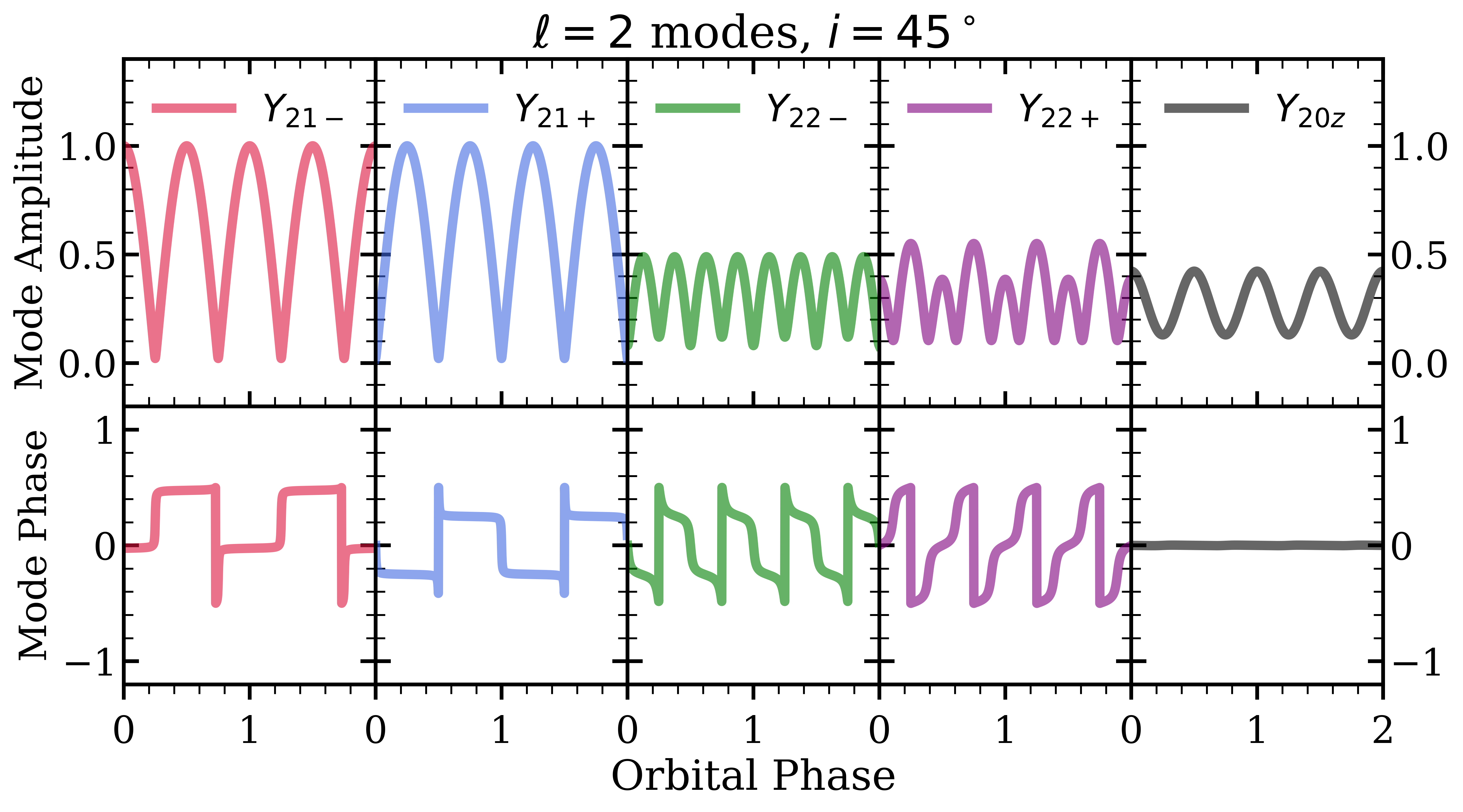}
\caption{\label{fig:l2amp} Same as Figure \ref{fig:l1amp} for the $\ell=2$ pulsation modes shown in Figure \ref{fig:l2lightcurve}. Note that phase -0.5 and 0.5 are identical, so transitions between these values are not actual phase changes. The $Y_{21\pm}$ modes have two phase jumps per orbit, the $Y_{22\pm}$ modes have four phase jumps, and the $Y_{20z}$ mode has none.
}
\end{figure*}

\begin{figure}
\includegraphics[scale=0.37]{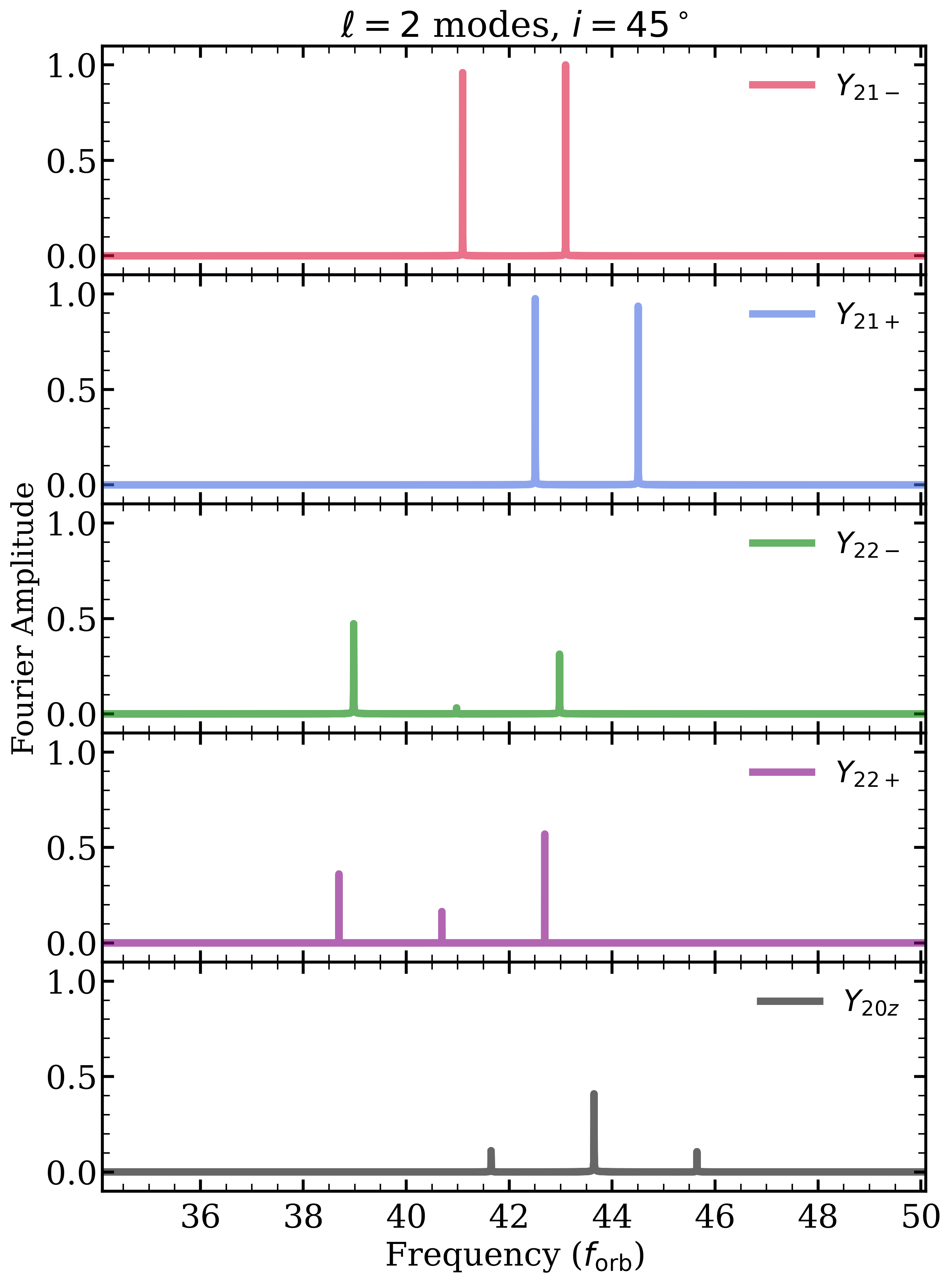}
\caption{\label{fig:l2fourier} Same as Figure \ref{fig:l1fourier} for the $\ell=2$ pulsation modes shown in Figures \ref{fig:l2lightcurve} and \ref{fig:l2amp}. The $Y_{22-}$, $Y_{22+}$, and $Y_{20z}$ modes can form triplets with asymmetric amplitudes, somewhat different from the idealized modes of Section \ref{sec:quad}.}
\end{figure}

The second solution also has frequency
\begin{equation}
    \omega^2 \simeq \omega_\alpha^2 + \delta \omega_{22}^2
\end{equation}
and eigenvector ${\bf a} \simeq [1, 0, 0, 0, 1]$. The spatial dependence is
\begin{equation}
    \delta F \propto \big[Y_{22}(\theta,\phi) + Y_{2-2}(\theta,\phi) \big] e^{- i \omega_0 t} \, 
\end{equation}
which reduces to
\begin{align}
\label{eq:dfb}
    \delta F &\propto \sin^2 \theta (\cos^2 \phi - \sin^2 \phi) \cos (\omega_0 t) \nonumber \\
    &\propto (x^2 - y^2) \cos (\omega_0 t) \, .
\end{align}
This mode also has an $\ell=m=2$ pattern that is wrapped around the $z$-axis, now with maxima along with the tidal axis, and it is a standing mode that does not propagate around the $z$-axis. We refer to it as the $Y_{22+}$ mode. The only difference between the $Y_{22-}$ and $Y_{22+}$ eigenfunctions is their phase relative to the $x$ and $y$-axes.

The third solution has
\begin{equation}
    \omega^2 \simeq \omega_\alpha^2 + \delta \omega_{00}^2
\end{equation}
and eigenvector ${\bf a} \simeq [0, 0, 1, 0, 0]$. Hence, this mode is essentially the normal $m=0$ mode, so we refer to it as the $Y_{20z}$ mode.




\subsubsection{Quadrupole mode visibility and amplitude modulation}

Figures \ref{fig:l2lightcurve}, \ref{fig:l2amp}, and \ref{fig:l2fourier} show the light curves, amplitude/phase modulation, and power spectra associated with triaxial $\ell=2$ modes. They show numerical solutions for the tidally coupled modes for the $\delta$\,Scuti model from Section \ref{sec:application}, but now for the $\ell=2$ and $n_{\rm pg} = 8$ p modes. The $Y_{21-}$ and $Y_{21+}$ modes display similar behavior to the $Y_{10x}$ and $Y_{10y}$ modes discussed above, with their amplitude modulated twice per orbit with phase jumps of 0.5 cycles. Similarly, their power spectra have two peaks spaced by twice the orbital frequency. This means that it will be difficult to observationally distinguish between $Y_{21-}$ and $Y_{10x}$ modes, or between $Y_{21+}$ and $Y_{10y}$ modes. 

The $Y_{22-}$ and $Y_{22+}$ modes have their amplitude modulated four times per orbit, with four jumps in phase by 0.5. Consequently their power spectra are two amplitude peaks separated by exactly four times the orbital frequency. They differ only in phase: the $Y_{22-}$ mode has zero amplitude at the epochs of conjunction (e.g., at eclipses), while the $Y_{22+}$ mode has its maximum amplitude at conjunctions. 

Finally, the $Y_{20z}$ mode behaves the same way as a normal $Y_{20}$ mode. It exhibits no phase or amplitude modulation, and produces a single peak in the power spectrum, the same as the $Y_{10z}$ modes and radial modes. Therefore, mode identification for these modes is difficult, perhaps requiring forward modelling -- i.e., by matching observed and model frequencies for these modes. Eclipse mapping could be used to distinguish between different types of modes.


Our analytic results for quadrupole modes use approximations not strictly valid in systems with order unity mass ratios (see Section \ref{sec:quad}). As can be seen from the numerical results in Figures \ref{fig:l2lightcurve}, \ref{fig:l2amp}, and \ref{fig:l2fourier} , the $Y_{22-}$, $Y_{22+}$, and $Y_{20z}$ eigenfunctions are slightly mixed with each other, complicating the amplitude modulation and power spectra. This introduces substantial amplitude variation to the $Y_{20z}$ mode, and produces side lobes to its central peak in the amplitude spectrum. Similarly, the the $Y_{22-}$, $Y_{22+}$ modes may obtain a central peak in their amplitude spectra, and asymmetric amplitudes of their primary peaks spaced by four times the orbital frequency. Such asymmetric peaks are also observed in roAp stars due to competition between rotation and magnetic effects \cite{kurtz:86}, whereas here it is a competition between rotation and tidal effects.


\subsubsection{Origin of the Phase Jumps}

In terms of their azimuth ($\phi$) and time dependence, we can rewrite equations \ref{eq:xi+2}, \ref{eq:xi-2}, \ref{eq:dfa}, and \ref{eq:dfb} as
\begin{eqnarray}
Y_{21+} \propto \sin \phi \cos(\omega_+ t) \\
Y_{21-} \propto \cos \phi \cos(\omega_- t) \\
Y_{22+} \propto \cos 2\phi \cos(\omega_0 t) \\
Y_{22-} \propto \sin 2\phi \cos(\omega_0 t) \, .
\end{eqnarray}
In these forms, it is apparent that the observer sees a pure sinsoidal frequency multiplied by a sinusoid of phase $\phi$ (in the case of $Y_{21\pm}$) or $2\phi$ (in the case of $Y_{22\pm}$).  The observer's view passes through these azimuths as the orbit revolves.  For each zero crossing of the sinusoid with $\phi$, the phase of the pulsation is reversed---twice or four times per orbit for the $\phi$ and $2\phi$ terms, respectively. It is exactly these phase reversals that lead to the splitting of the modes (by $\pm \nu_{\rm orb}$ or $\pm 2 \nu_{\rm orb}$) in the observer's frame in which the binary is orbiting (and the pulsating star is rotating).  Analogous phase shifts (two per orbit) are seen by the observer for the $Y_{10x}$ and $Y_{10y}$ modes. If these modes were traveling waves (i.e., ordinary $Y_{11}$ or $Y_{1-1}$ modes), as opposed to the standing waves they actually are, there would be no phase jumps observed around the orbit and no splitting of the modes.

\section{Tidal coupling approximation}
\label{sec:approx}

The analysis above is thorough but does not illuminate the most important terms in the tidal coupling calculation. Here we provide simple and accurate expressions for the tidal coupling coefficients (the values of $\delta \omega_{m m'}^2$ in equation \ref{eq:matrixcoup}) for $\ell=1$ p~modes and g~modes in the asymptotic limit.

\subsection{p~modes}

We begin with the value of $\delta T_{mm'}$ in equation \ref{eq:dt}. For p~modes, the dominant terms are those containing the radial displacement $\xi_r$ and its radial derivatives. We thus find
\begin{equation}
\label{eq:dtap}
    \delta T_{mm'} \simeq - X_{\ell 2 \ell'}^{m m_t m'} \int^{R}_0 \frac{2}{3} \varepsilon \rho r^2 (\eta + 3) \xi_r^2 dr \, .
\end{equation}
The tidal distortion is largest in the outer part of the star where the density is small so that $\eta \simeq 3$. Using equation \ref{eq:epsval}, we have 
\begin{equation}
\label{eq:dtap2}
    \delta T_{mm'} \simeq - 3 X_{\ell 2 \ell'}^{m m_t m'} \frac{M_c}{M} \frac{R^3}{a^3} \int^{R}_0 \rho r^2 \frac{r^3}{R^3} \xi_r^2 dr \, .
\end{equation}

Performing the same exercise for the $\delta V$, the dominant terms for p~modes are
\begin{align}
\label{eq:dvap}
    \delta V_{mm'} &\simeq X_{\ell 2 \ell'}^{m m_t m'} \int^{R}_0 \frac{2}{3} \varepsilon \rho r^2 c_s^2 \nonumber \\
    &\times \bigg[ -2 \bigg(\frac{d \xi_r}{dr}\bigg)^2 + (\eta +1) \bigg(\frac{d \xi_r}{dr}\bigg)^2 \bigg] dr \, .
\end{align}
For p~modes, $(d \xi_r/dr)^2 \simeq (\omega^2/c_s^2) \xi_r^2$. Using the same approximations above, we have 
\begin{align}
\label{eq:dvap2}
    \delta V_{mm'} &\simeq X_{\ell 2 \ell'}^{m m_t m'} \frac{M_c}{M} \frac{R^3}{a^3} \omega_\alpha^2 \int^{R}_0 \rho r^2 \frac{r^3}{R^3} \xi_r^2 dr \nonumber \\
    &\simeq \frac{-\omega_\alpha^2}{3} \delta T_{m m'} \, .
\end{align}
Denoting the integral (which has dimensionless units for our choice of normalization) as $\mathcal{P}$, we have
\begin{equation}
    \delta V_{m m'} - \omega_\alpha^2 \delta T_{m m'} \simeq 4 X_{\ell 2 \ell'}^{m m_t m'} \frac{M_c}{M} \frac{R^3}{a^3} \omega_\alpha^2  \, \mathcal{P} \, .
\end{equation}
For the low-order p~modes of our $\delta$\,Scuti model discussed below, we find $\mathcal{P} \sim 0.5$, with little dependence on radial order, but this value may change somewhat for stars with different structures.

Summing over each source of ellipticity (tidal distortion and centrifugal distortion), and using $X_{1 2 1}^{1 0 1} = -1/5$, $X_{1 2 1}^{0 0 0} = 2/5$, $\sqrt{\frac{3}{2}} X_{1 2 1}^{1 2 -1} = -3/5$, we find the perturbed matrix elements to be
\begin{align}
\label{eq:matrixcoupap}
\delta \omega_{m m'}^2 &\simeq \frac{4}{5} \frac{M_c}{M} \frac{R^3}{a^3} \omega_\alpha^2  \, \mathcal{P} \nonumber \\
&\times
\begin{bmatrix}
-1 - \frac{2(M+M_c)}{3 M_c} & 0 & 3 \\
0 & 2 + \frac{4(M+M_c)}{3 M_c} & 0 \\
3 & 0 & -1 - \frac{2(M+M_c)}{3 M_c}
\end{bmatrix}
\end{align}
For high-order p~modes, the additional centrifugal and Coriolis terms are negligible relative to these for stars in close binaries. 

The $m=0$ mode is uncoupled to the others and forms the $Y_{10z}$ mode. From equation \ref{eq:matrixcoupap}, we can see that its frequency perturbation will be
\begin{equation}
\label{eq:omz}
    \frac{\delta \omega_{z}}{\omega_\alpha} \approx \frac{4}{5} \frac{M_c}{M} \frac{R^3}{a^3} \mathcal{P} \bigg(1 + \frac{2(M+M_c)}{3 M_c} \bigg) \, .
\end{equation}
Note the $Y_{10z}$ mode always has a positive frequency perturbation because the tidal/centrifugal ellipticity always shortens the height of the star along the $z$-axis, so it takes less time for pressure waves to propagate through the star  in the $z$-direction. 

The $m=\pm1$ modes are coupled to each other and from equation \ref{eq:matrixcoupap} their frequency changes are
\begin{equation}
    \frac{\delta \omega_{\pm}}{\omega_\alpha} \approx \frac{2}{5} \frac{M_c}{M} \frac{R^3}{a^3} \mathcal{P} \bigg[-\bigg( 1 + \frac{2(M+M_c)}{3 M_c}\bigg) \pm 3 \bigg] \, .
\end{equation}
The negative root is the $Y_{10x}$ mode where we find
\begin{equation}
\label{eq:omy}
    \frac{\delta \omega_x}{\omega_\alpha} \approx - \frac{2}{5} \frac{M_c}{M} \frac{R^3}{a^3} \mathcal{P} \bigg[4 + \frac{2(M+M_c)}{3 M_c} \bigg] \, .
\end{equation}
The $Y_{10x}$ mode always has a negative frequency perturbation because the tidal/centrifugal ellipticity always increases the length of the star along the $x$-axis, so it takes more time for pressure waves to propagate through the star in this direction. The positive root is the $Y_{10y}$ mode where we find
\begin{equation}
\label{eq:omx}
    \frac{\delta \omega_y}{\omega_\alpha} \approx \frac{2}{5} \frac{M_c}{M} \frac{R^3}{a^3} \mathcal{P} \bigg[2 - \frac{2(M+M_c)}{3 M_c} \bigg] \, .
\end{equation}
This frequency perturbation can be positive or negative, depending on the mass ratio, because of the competing effects of tidal and centrifugal distortion in determining the length of the star in the $y$-direction.

We can see that the tidal frequency perturbation is comparable to the dimensionless tidal distortion $\epsilon = (M_c/M) (R/a)^3$, for mass ratios of order unity. Note that for vanishing companion masses ($M_c \rightarrow 0$) the second term in brackets dominates, which is the term arising from the centrifugal force. In this limit, the star is nearly axisymmetric about the rotation axis, and the $Y_{10x}$ and $Y_{10y}$ modes have almost the same frequency. In the opposite limit where $M_c \gg M$ (but with larger $a$ such that $\epsilon$ is unchanged), the centrifugal term becomes negligible, and the star's distortion is nearly axisymmetric about the tidal axis. In this limit, the $Y_{10y}$ and $Y_{10z}$ modes have the same frequency. For mass ratios of order unity as we expect in most stellar binaries, the three modes have similar frequency perturbations, and their splitting is not typically symmetric. We demonstrate that the approximations of equations \ref{eq:omz}-\ref{eq:omx} are very good in Section \ref{sec:application}. Stars in close binaries with large values of $\epsilon$ will be asymmetric across the $y-z$ plane due to $\ell_t=3$ components of the tidal potential, which are not included in our calculations and will also induce frequency perturbations.

\subsection{g~modes}

For g~modes, the dominant terms in the tidal coupling integrals are those containing the horizontal displacement $\xi_\perp$. Using the same procedure as above, we find
\begin{equation}
\label{eq:dtapg}
    \delta T_{mm'} \simeq \frac{1}{2} X_{\ell 2 \ell'}^{m m_t m'} \frac{M_c}{M} \frac{R^3}{a^3} \int^{R}_0 \rho r^2 \frac{r^3}{R^3} (\eta+3) \xi_\perp^2 dr \, .
\end{equation}
Since g~modes are mostly confined to the near-core regions of main sequence stars, we cannot use $\eta \simeq 3$ as we did above, but our models typically have $\eta$ of order unity in the g~mode cavity.

For $\delta V$, the dominant terms for $\ell=1$ g~modes add up to 
\begin{equation}
\label{eq:dvapg}
    \delta V_{mm'} \simeq X_{\ell 2 \ell'}^{m m_t m'} \frac{M_c}{M} \frac{R^3}{a^3} \int^{R}_0 \rho r^2 L_1^2 \frac{r^3}{R^3} (2 \eta + 5) \xi_\perp^2 dr \, ,
\end{equation}
where $L_1^2 = 2 c_s^2/r^2$ is the squared Lamb frequency. Since g~modes typically have $\omega^2 \ll L_1^2$, then $\delta V \gg \delta T$, and we can ignore the kinetic energy terms. Hence, the tidal component of the perturbed matrix elements for g~modes are 
\begin{equation}
    \delta V_{m m'} - \omega_\alpha^2 \delta T_{m m'} \simeq X_{\ell 2 \ell'}^{m m_t m'} \frac{M_c}{M} \frac{R^3}{a^3} \, \mathcal{G} \, ,
\end{equation}
where $\mathcal{G}$ is the integral in equation \ref{eq:dvapg}.

Similar to p~modes, the tidal coupling matrix is thus 
\begin{align}
\label{eq:matrixcoupgap}
\delta \omega_{m m'}^2 &\simeq \frac{1}{5} \frac{M_c}{M} \frac{R^3}{a^3}  \, \mathcal{G} \nonumber \\
&\times
\begin{bmatrix}
-1 - \frac{2(M+M_c)}{3 M_c} & 0 & 3 \\
0 & 2 + \frac{4(M+M_c)}{3 M_c} & 0 \\
3 & 0 & -1 - \frac{2(M+M_c)}{3 M_c}
\end{bmatrix}
\end{align}
In Section \ref{sec:discussion}, we show that these terms are typically smaller than the Coriolis terms for g~modes in main sequence stars, so we do not usually expect g~modes to be tidally tilted.


\section{Application to $\delta$\,Scuti pulsators}
\label{sec:application}

In this section, we apply the theory developed above to pulsations of $\delta$\,Scuti pulsators in close binaries, which comprise the majority of tidally tilted pulsators observed thus far. However, it could apply equally well to other types of p~mode pulsators, such as $\beta$\,Cephei or sdBV stars. 

We begin by constructing a stellar model using the MESA stellar evolution code \citep{paxton:11,paxton:13,paxton:15,paxton:18,paxton:19}. Our chosen model is similar to that of the triaxial $\delta$\,Scuti pulsator TIC 184743498 \citep{zhang:24}. It has a mass $M=1.74 \, {\rm M}_\odot$, radius $R=2.23 \, {\rm R}_\odot$, metallicity $Z=0.017$, surface temperature $T_{\rm eff} = 7427 \, {\rm K}$, and is roughly halfway through core hydrogen burning. We use the GYRE pulsation code \citep{townsend:13} to compute the non-adiabatic $\ell=1$ pulsation modes of our stellar model, including low-order g~modes and p~modes. 

\begin{figure}
\includegraphics[scale=0.32]{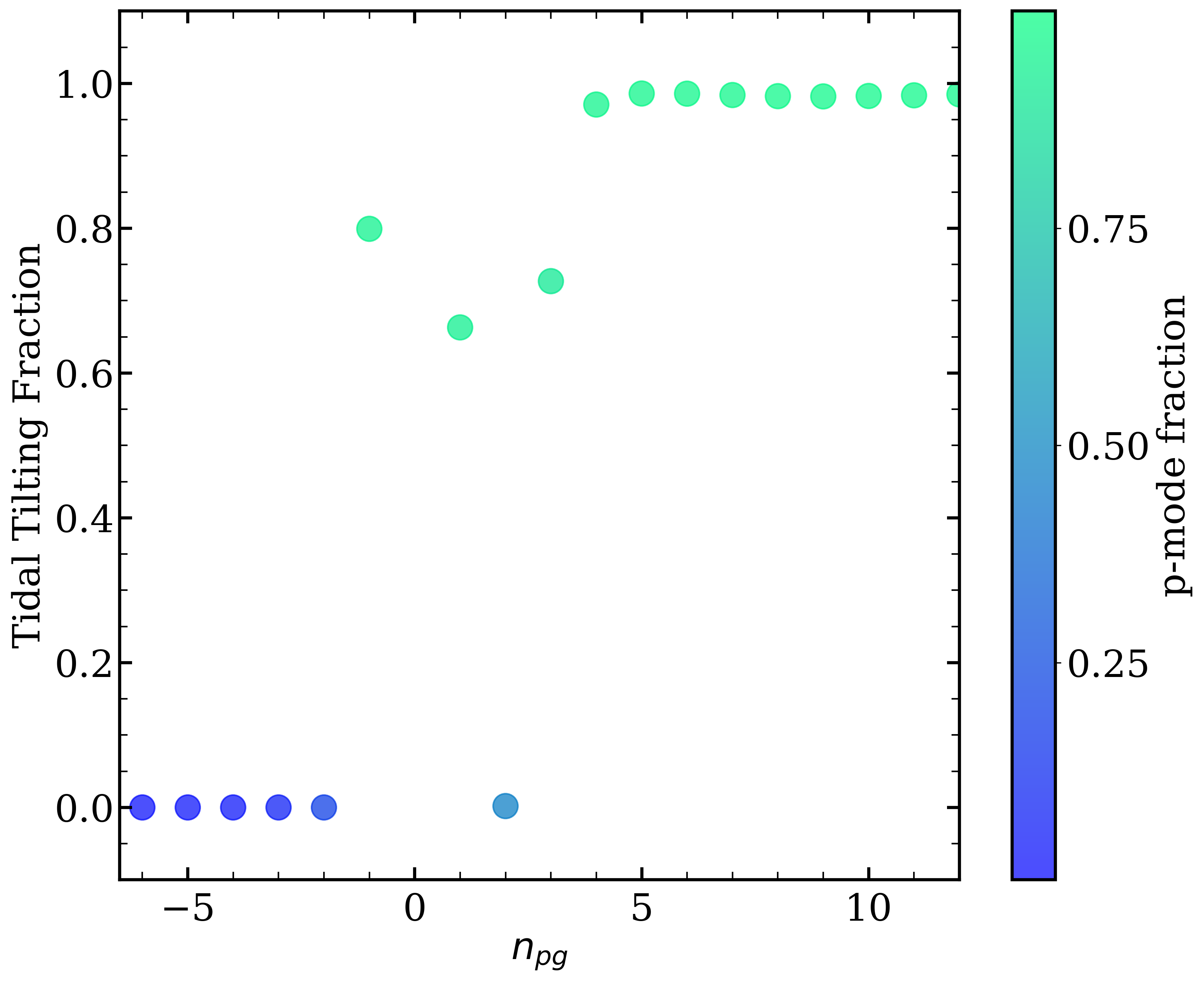}
\caption{\label{fig:tidaltilt} The tidal tilting fraction of dipole modes in our $\delta$\,Scuti model at an orbital period of $P_{\rm orb} = 3 \,{\rm d}$, as a function of mode radial order $n_{\rm pg}$. The tidal tilting fraction is defined to be unity for tidally aligned modes, while it is zero for modes aligned with the spin axis. Color indicates the p~mode contribution to the total mode energy. We see that p~modes tend to be tidally tilted, while g~modes remain aligned with the star's spin axis.
}
\end{figure}

For each oscillation mode, we compute the tidal overlap integrals $T_{\rm int}$ and $V_{\rm int}$ defined in equations \ref{eq:dt} and \ref{eq:dV}, as well as the Coriolis term $W$ from equation \ref{eq:wmat}, and the centrifugal perturbations $\delta V_{\rm cen}$ from equation \ref{eq:dvcen}. We next compute the Wigner coefficients and tidal distortion required to compute the $\delta \omega^2_{mm'}$ terms in equation \ref{eq:matrixcoup}. In what follows, we use a companion mass of $M_c = 1.3 \, {\rm M}_\odot$ at orbital periods $1 \, {\rm d} \leq P_{\rm orb} \leq 3 \, {\rm d}$, short enough for the system to be tidally circularized and synchronized \citep{bashi:23} but long enough that the star is not close to filling its Roche lobe.

We solve the matrix equation \ref{eq:matrixcoup} for the frequencies $\omega$ and eigenvectors ${\bf a}$ of the $\ell=1$ tidally tilted modes of our model. To quantify whether the modes remain aligned with the rotation axis, or whether they become triaxial pulsations, we define the tidal tilting fraction for the $m=\pm1$ modes as
\begin{equation}
    Z_{\rm tilt} = 1 - |a_{-1}^2 - a_{1}^2 |
\end{equation}
where $a_{m}$ is the contribution of the $Y_{1m}$ spherical harmonic to each of the new tidally tilted modes. This is defined such that $Z_{\rm tilt} = 0$ in the absence of tidal distortion, and $Z_{\rm tilt}=1$ for triaxial modes with $a_{-1}^2 = a_{1}^2$.

Figure \ref{fig:tidaltilt} shows the tidal tilting fraction as a function of mode radial order $n_{\rm pg}$ at an orbital period $P_{\rm orb} = 3 \, {\rm d}$. The radial order $n_{\rm pg} < 0$ for predominantly g~modes and $n_{\rm pg} > 0$ for predominantly p~modes. We can see that g~modes with $n_{\rm pg}\leq -2$ are almost fully aligned with the rotation axis ($Z_{\rm tilt} \simeq 0)$ while p~modes with $n_{\rm pg} \geq 4$ are almost fully tidally tilted ($Z_{\rm tilt} \simeq 1)$. The modes in between are mixed modes that have some p~mode character in the envelope and some g~mode character in the core. For these modes, the tidal tilting fraction can have intermediate values. This means such modes have significant (but not equal) contributions from both $m=-1$ and $m=1$ components. They would produce doublets of unequal amplitude split by twice the orbital frequency in a power spectrum. We find very similar results at somewhat shorter or longer periods, except that the tidal tilting fraction for the $n_{\rm pg} \sim 0$ modes becomes slightly larger or smaller, respectively. 

\begin{figure}
\includegraphics[scale=0.38]{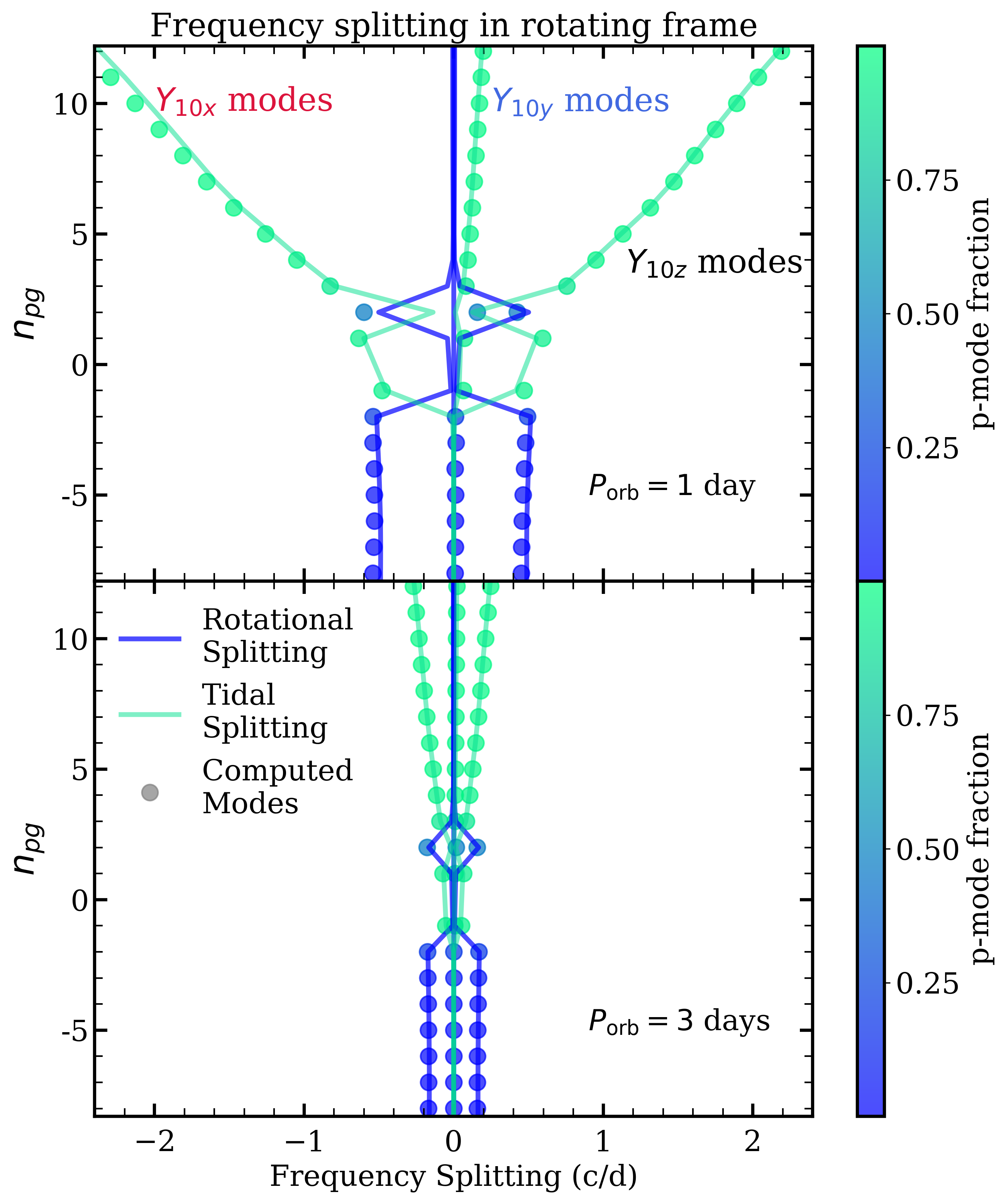}
\caption{\label{fig:tidalsplitper} The combined tidal and rotational frequency perturbation (measured in the star's rotating frame) to modes in our $\delta$\,Scuti model at an orbital period of $P_{\rm orb} = 1 \,{\rm d}$ (top panel) and $P_{\rm orb} = 3 \,{\rm d}$ (bottom panel). The mode radial order $n_{\rm pg}$ is on the y-axis. The blue lines indicate the expected frequency splitting due to the Coriolis force, while green lines indicate the expected frequency splitting due to tides (Equations \ref{eq:omz}-\ref{eq:omx}). The g~modes are perturbed primarily by the Coriolis force, while p~modes are perturbed primarily by the tidal distortion.
}
\end{figure}

The color of the points in Figure \ref{fig:tidaltilt} is the p~mode fraction of the modes, which we define as 
\begin{equation}
    p_{\rm frac} = \frac{ \int \rho r^2 |\xi_r|^2 dr}{ \int \rho r^2 |\bxi|^2 dr} \, .
\end{equation}
Similarly, the g~mode fraction is 
\begin{equation}
    g_{\rm frac} = \frac{ \int \rho r^2 |\xi_\perp|^2 dr}{ \int \rho r^2 |\bxi|^2 dr} = 1 - p_{\rm frac} \, .
\end{equation}
Modes that are p-dominated tend to be tidally tilted, while g-dominated modes are rotationally aligned. However, low-order p~modes can still be rotationally aligned depending on the amplitude of the tidal distortion, which is smaller for longer period binaries. Our models predict that high-order p~modes are more likely to be tidally tilted than low-order p~modes (see Section \ref{sec:when}), which can be tested with future observations.

The mode frequency perturbations also demonstrate the sensitivity of p~modes to tidal distortion. Figure \ref{fig:tidalsplitper} shows the frequency perturbations of modes relative to that of a non-rotating and non-tidally distorted star, for different radial orders $n_{\rm pg}$. The blue lines indicate the approximate frequency splitting expected for due to the Coriolis force, 
\begin{align}
    \delta \omega &= - m \Omega C_\alpha \\
    &\simeq - \frac{m}{\ell(\ell+1)} \Omega \quad {\rm for \, g~modes} \nonumber \\
    &\simeq 0 \quad {\rm for \, p~modes} \, , \nonumber
\end{align}
where $C_\alpha$ is the Ledoux constant (Equation \ref{eq:wmat}). The green lines indicate the approximate splitting expected for p~modes due purely to tidal distortion (equations \ref{eq:omz}-\ref{eq:omx}). Note that while the tidal splitting can be small for g modes, and the rotational splitting can be small for p modes, they are non-zero. This allows us to distinguish between rotational splitting (which produces frequencies split by nearly the orbital/spin frequency) and tidal amplitude modulation (which produces peaks split by exact integer multiples of the orbital frequency.

\begin{figure}
\includegraphics[scale=0.38]{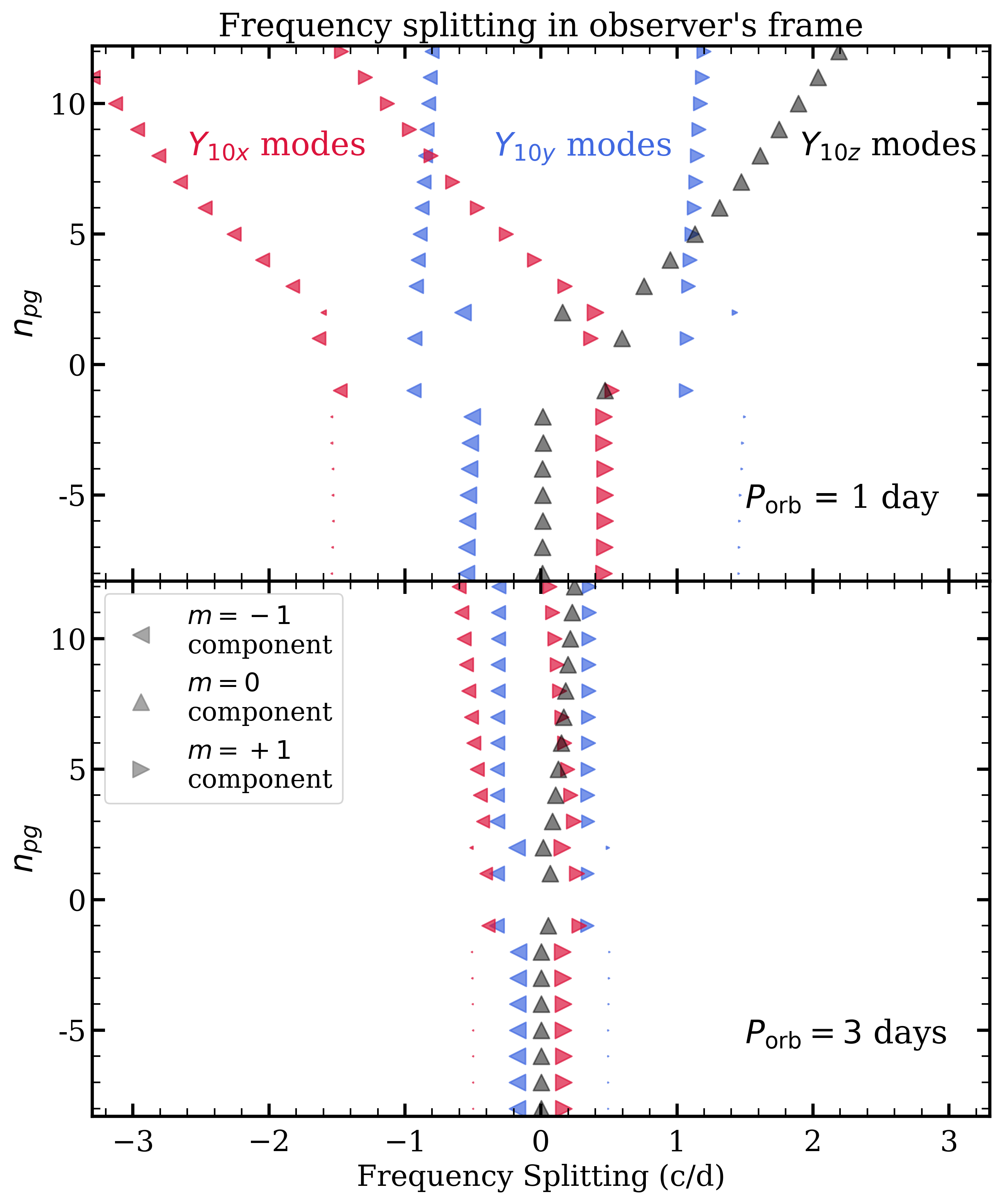}
\caption{\label{fig:tidalsplitperobs} Combined tidal and rotational frequency perturbations for modes of our $\delta$\,Scuti model, but now measured in the observer's frame. The triangle symbols indicate the contribution from different $m$ values to each mode, whose color is arbitrary. The g~modes form singlets, with different $m$ components spaced by roughly half the orbital frequency, similar to single stars. The $Y_{10z}$ modes are singlets with only an $m=0$ component, but the $Y_{10x}$ and $Y_{10y}$ modes have $m=\pm1$ components, forming doublets split by exactly twice the orbital frequency. The p~modes perturbed to lower frequency are $Y_{10x}$ mode doublets, while those perturbed to higher frequencies are $Y_{10z}$ mode singlets, and the $Y_{10y}$ mode doublets are nearly unperturbed.
}
\end{figure}

It is clear that high-order p~modes have frequency perturbations dominated by tidal distortion, whereas high-order g~modes have frequency perturbations dominated by the Coriolis force, at both $P_{\rm orb} = 1 \, {\rm d}$ and $P_{\rm orb} = 3 \, {\rm d}$. Note that the tidal frequency perturbation is nearly a constant fraction of the unperturbed frequency, meaning that the frequency perturbation is larger for higher frequency p~modes. The tidal frequency perturbation can be smaller than the rotation rate for low-order p~modes at $P_{\rm orb} = 3 \, {\rm d}$, despite the fact that those p~modes are nearly fully tidally tilted. At $P_{\rm orb}=1 \, {\rm d}$, the tidal frequency perturbation to p~modes tends to be much larger than the rotational splitting of g~modes, especially for high-frequency p~modes. 

Figure \ref{fig:tidalsplitperobs} shows the corresponding observed mode frequencies of our stellar model. Here, the $m=\pm1$ modes produce a doublet in the observed frequencies due to tidal tilting, and the size of the symbol indicates the relative amplitude of each component of the doublet. Each g~mode appears as a singlet because they are not tidally tilted, so each radial order of dipole modes produces triplets, just as in normal pulsators. In contrast, the $Y_{10x}$ and $Y_{10y}$ p~modes appear as equal-amplitude doublets split by twice the orbital frequency. Including the singlets produced by $Y_{10z}$ modes, each radial order of dipole p~modes thus produces five peaks in the power spectrum. As shown above, the $Y_{10x}$ modes are shifted to lower frequencies, while $Y_{10z}$ modes are shifted to higher frequencies, and $Y_{10y}$ modes are in between. The splitting by 2$\times$ orbital frequency can be larger or smaller than the tidal frequency perturbation, meaning that the $Y_{10x}$ mode and $Y_{10y}$ mode doublets and $Y_{10z}$ mode singlets may overlap in frequency space, and will not typically produce neatly organized quintuplets.

\begin{figure}
\includegraphics[scale=0.38]{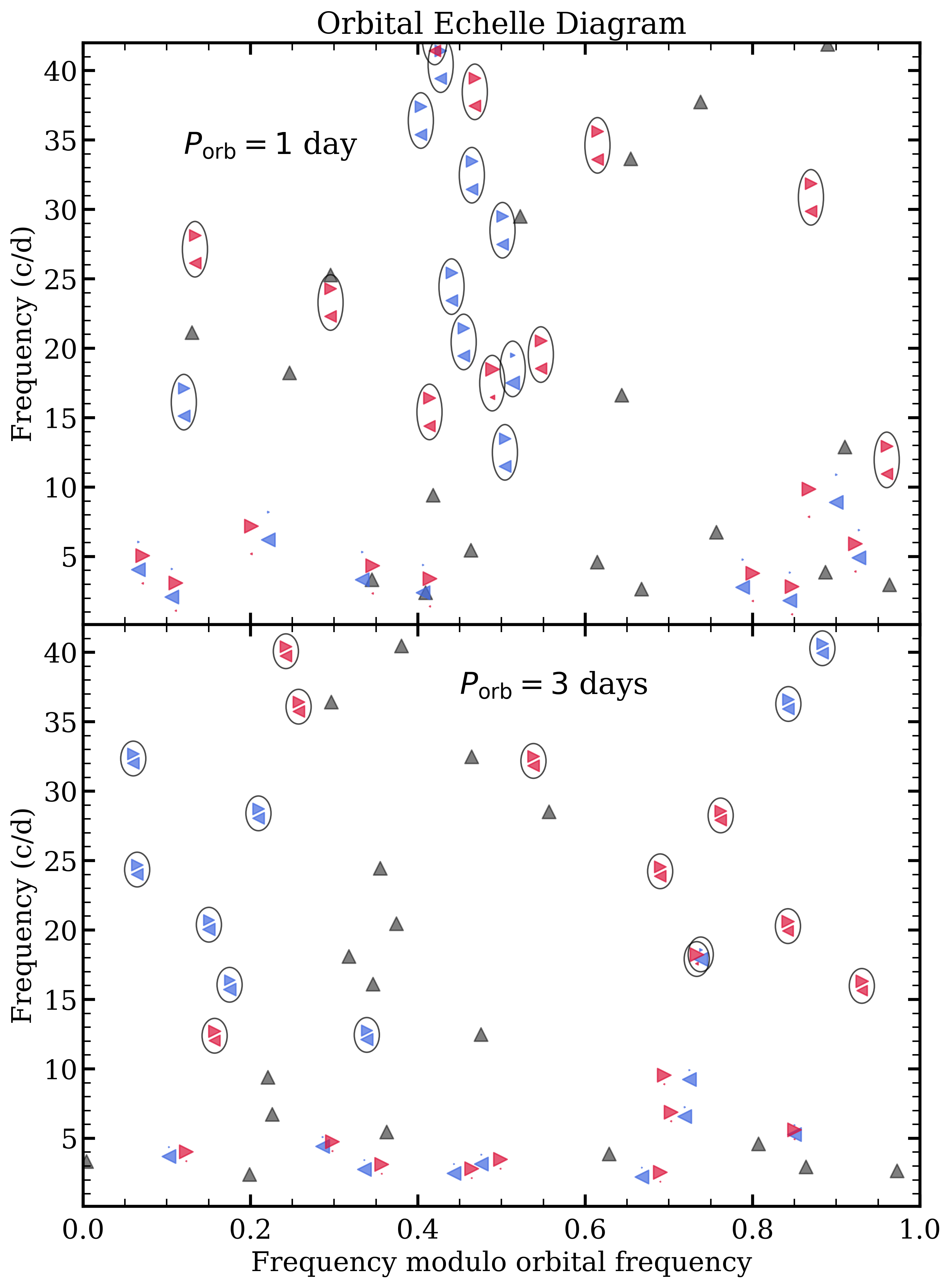}
\caption{\label{fig:tidalsplitperechorb} Echelle diagram for the dipole modes of our $\delta$\,Scuti model, as a function of the frequency modulus relative to the orbital frequency. Symbol size and color is the same as Figure \ref{fig:tidalsplitperobs}. Each tidally tilted $Y_{10x}$ and $Y_{10y}$ mode forms a vertically aligned doublet, separated by twice the orbital frequency, which is surrounded by an ellipse for clarity. The $Y_{10z}$ modes form singlets. Note the doublets are always formed by $m=-1$ and $m=1$ components, while the $Y_{10z}$ modes are pure $m=0$ modes. Finding doublets on these Echelle diagrams can be used to identify tidally tilted pulsators, but there will usually be contamination from singlets due to $Y_{10z}$ modes or radial modes.
}
\end{figure}

To find and properly interpret tidally tilted pulsations, it may be most useful to make Echelle diagrams modulo the orbital frequency. Figure \ref{fig:tidalsplitperechorb} shows the predicted Echelle diagram of our $\delta$\,Scuti model. Tidally tilted modes show up as equal-amplitude vertically aligned doublets. Spin-aligned g~modes or $Y_{10z}$ modes appear as singlets. Because the orbital frequency has no relation to the star's large frequency spacing, $\Delta \nu$, the frequency moduli in Figure \ref{fig:tidalsplitperechorb} are scattered with no apparent pattern. This is qualitatively consistent with the patterns seen in the triaxial pulsators studied by \cite{zhang:24} and \cite{jayaraman:24}.

\begin{figure}
\includegraphics[scale=0.38]{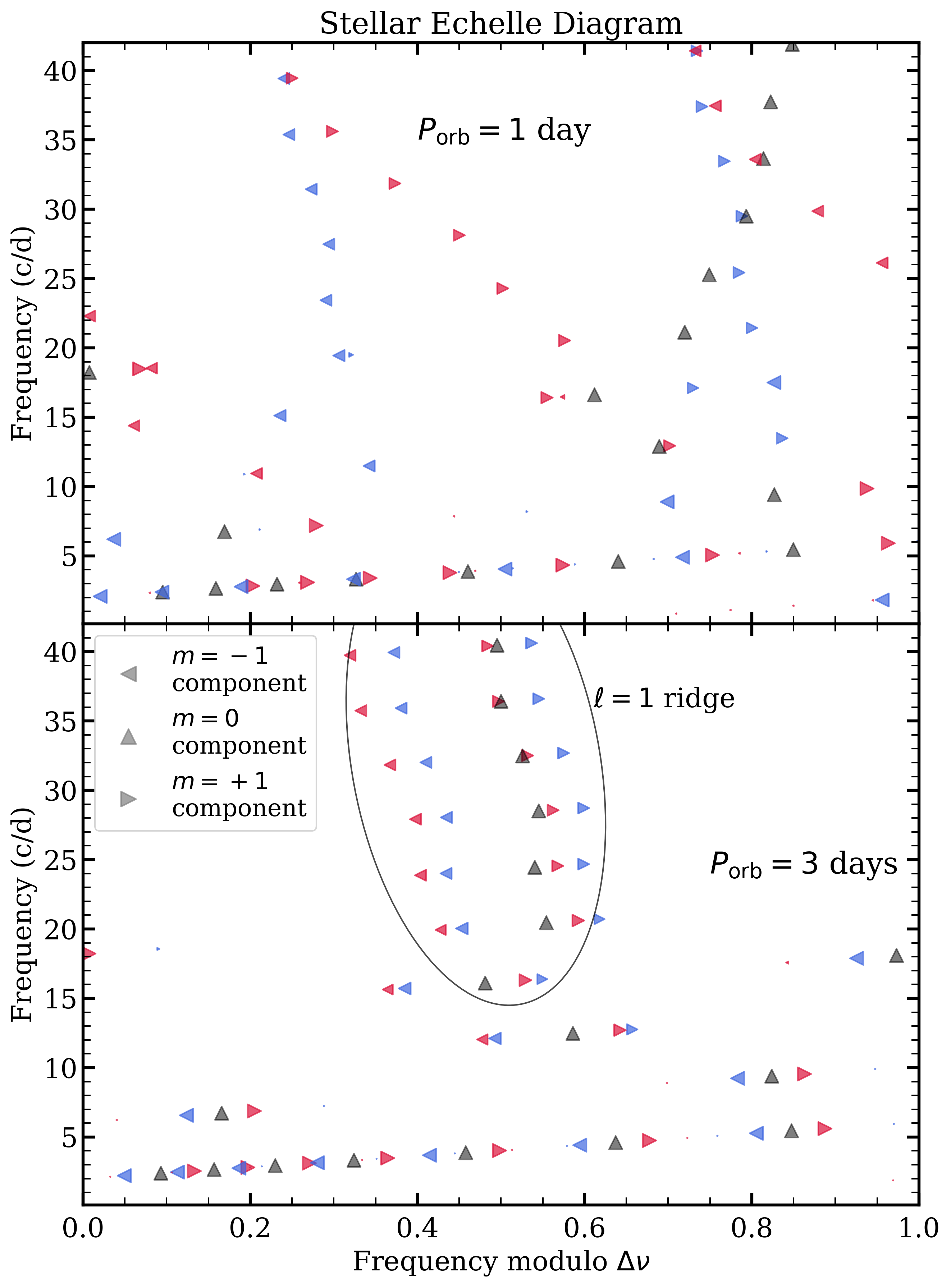}
\caption{\label{fig:tidalsplitperechstar} Echelle diagram for our $\delta$\,Scuti model, as a function of the frequency modulus relative to the large frequency spacing $\Delta \nu$ of a single non-rotating star. At $P_{\rm orb} = 3 \, {\rm d}$ (bottom panel), the tidal frequency perturbations are small enough that the p~modes remain confined to a vertical ridge in the Echelle diagram (surrounded by an ellipse for clarity), although this ridge is broadened by the splitting of the $Y_{10x}$ and $Y_{10y}$ modes by twice the orbital frequency. At $P_{\rm orb} = 1 \, {\rm d}$ (top panel), the tidal frequency perturbations are comparable to $\Delta \nu$ such that the modes are scattered across a wide range of frequency moduli. 
}
\end{figure}

One may wonder whether the star's large frequency spacing,  $\Delta \nu$, can be determined using an Echelle diagram in tidally tilted pulsators, in the same manner it can be determined for single and slowly rotating stars. Figure \ref{fig:tidalsplitperechstar} shows the Echelle diagram of our model using the star's asymptotic non-rotating value of $\Delta \nu$. At longer orbital periods ($P_{\rm orb} = 3 \, {\rm d}$), the tidal frequency perturbations remain small relative to $\Delta \nu$, so p~modes still show up as a vertical ridge in the Echelle diagram, although it has been broadened by the 2$\times$ the orbital frequency due to tidal tilting. At short periods ($P_{\rm orb} = 1 \, {\rm d}$), the tidal frequency perturbations become comparable to or larger than $\Delta \nu$, causing the modes to have a wide range of frequency moduli, so that $\Delta \nu$ will be difficult to measure.

\section{Discussion}
\label{sec:discussion}

\subsection{When will modes be tidally tilted?}
\label{sec:when}

\subsubsection{p~modes}

It is important to understand when modes will be triaxial pulsations that are tidally tilted, and when they will remain aligned with the rotation axis. As discussed in Section \ref{sec:dipole}, dipole modes will be in the tidally tilted limit when the off diagonal tidal coupling elements $\delta \omega_{1-1}^2$ are larger than the difference between the on-diagonal frequency shifts, $|\delta \omega_{11}^2 - \delta \omega_{-1-1}^2|$. For p~modes, this occurs when 
\begin{equation}
    \frac{12}{5} \frac{M_c}{M} \frac{R^3}{a^3} \omega_\alpha^2  \, \mathcal{P} \gtrsim 4 \omega_\alpha \Omega W_{\rm int} \, .
\end{equation}
This gives fractional radii larger than
\begin{equation}
\label{eq:ratilt}
    \frac{R}{a} \gtrsim \bigg( \frac{5}{3} \frac{M}{M_c} \frac{\Omega}{\omega_\alpha} \frac{W_{\rm int}}{\mathcal{P}} \bigg)^{1/3} \, 
\end{equation}
in order to be tidally tilted. In our $\delta$\,Scuti model, we find $\mathcal{P} \approx 0.5$, and the Ledoux constant is $C_{nl} = W_{\rm int}  \approx 0.005$ for high-frequency p~modes, without much dependence on radial order.

If the star is tidally synchronized, then $\Omega = \sqrt{G (M+M_c)/a^3}$. In this case, we find that p~modes will be tidally tilted for any orbital periods $P_{\rm orb} \lesssim 100 \, {\rm d}$, given $M=M_c=1.5 \, {\rm M}_\odot$, $R=1.5 \, {\rm R}_\odot$, and radial orders $n_{\rm pg}=10$. However, real binaries of this type are only observed to be tidally circularized and synchronized at periods less than $P_{\rm orb} \lesssim 3 \, {\rm d}$ \citep{bashi:23}, meaning that our model predicts p~modes to be tidally tilted for any $\delta$\,Scuti pulsator in a tidally synchronized binary.

If tidal tilting is not observed for tidally synchronized pulsators, it would indicate a shortcoming of our model, although it is not obvious what physical effect would prevent tidal tilting from occurring. The presence of a few modes without orbital modulation would not be sufficient to rule out tidal tilting, since such modes could be radial modes, $Y_{10z}$ modes or $Y_{20z}$ modes. Low-order p modes are also not expected to be tidally tilted to the same degree as high-order modes. Future work should closely examine whether tidal tilting is occurring in close binaries, and determine the fraction of pulsators in close binaries that show tidally tilted modes.


For binaries at periods longer than $P_{\rm orb} \approx 3 \, {\rm d}$, we must consider non-circular and non-synchronized rotation. If the star is not tidally synchronized, then equation \ref{eq:ratilt} can be rearranged to find that tidal tilting occurs when
\begin{equation}
\label{eq:porbtilt}
    P_{\rm orb} \lesssim \bigg( \frac{12 \uppi^2}{5} \frac{R^3}{G(M+M_c)} \frac{M_c}{M} \frac{f_\alpha}{f_{\rm rot}} \frac{\mathcal{P}}{W_{\rm int}} \bigg)^{1/2} \, 
\end{equation}
where $f_\alpha$ and $f_{\rm rot}$ are the linear pulsation frequency and rotation frequency, respectively. Using the same stellar parameters as above and $f_\alpha = 40 \, {\rm d}^{-1}$, $f_{\rm rot} = 1 \, {\rm d}^{-1}$, we find tidal tilting happens at periods  
\begin{equation}
\label{eq:porbtilt2}
    P_{\rm orb} \lesssim 6 \, {\rm d} \, \frac{f_\alpha}{40 \, {\rm d}^{-1}} \bigg( \frac{f_{\rm rot}}{1 \, {\rm d}^{-1}}\bigg)^{-1} \, . 
\end{equation}
Hence, we expect p~modes in typical $\delta$\,Scuti stars to be tidally tilted in binaries with periods of less than several days, but this number depends on both the mode frequency and the star's rotation frequency, as well as the stellar masses and radii.

\subsubsection{g~modes}

Tidal tilting could occur in g~mode pulsators such as $\gamma$\,Doradus or SPB stars. Performing the same analysis for g~modes, we expect tidal tilting to occur when
\begin{equation}
\label{eq:gtilt}
    \frac{3}{5} \frac{M_c}{M} \frac{R^3}{a^3} \, \mathcal{G} \gtrsim 4 \omega_\alpha \Omega W_{\rm int} \, .
\end{equation}
For $\ell=1$ g~modes, the Ledoux constant is $C_{nl} = W_{\rm int} \simeq 1/\ell(\ell+1) \simeq 1/2$. 
However, we caution that our perturbation theory breaks down for g~modes with $\omega_\alpha \lesssim 2 \Omega$, and g~modes are strongly altered by the Coriolis force into Hough modes (e.g., \citealt{bildsten:96,lee:97}). Since the Coriolis force is large in this regime, whereas the tidal distortion is always fairly small, we expect high-order g~modes ($n_{\rm pg} \lesssim -3$) to be aligned with the spin-axis of the star. 

For the lowest frequency g~modes applicable to our perturbation theory with $\omega_\alpha \sim 2 \Omega$, assuming a tidally synchronized star, equation \ref{eq:gtilt} can be rearranged to obtain the requirement for tidal tilting 
\begin{equation}
\label{eq:gtilt2}
     \mathcal{G} \gtrsim \frac{20}{3} \frac{M+M_c}{M_c} \frac{GM}{R^3} \, .
\end{equation}
However, we find that low-frequency g~modes typically have $\mathcal{G} \sim 0.1 GM/R^3$ in our $\delta$\,Scuti model, so equation \ref{eq:gtilt2} is never satisfied. Therefore we do not expect g~modes to be tidally tilted in close binaries.

A possible exception could occur in white dwarfs, where the g modes are trapped near the surface of the star where tidal effects are larger. This could increase the value of $\mathcal{G}$ and potentially allow for tidal tilting. This should be examined in future work.

\subsection{Tidal tilting vs. tidal trapping}

As mentioned above, tidal coupling also couples modes of different $\ell$, which is not accounted for in our isolated multiplet perturbation theory. Since the $\ell_t=2$ component of the tidal distortion is largest, coupling between modes differing by $\Delta \ell = 2$ will be important. In particular, coupling between $\ell=0$ and $\ell=2$ may significantly affect the geometries of these modes. Coupling tends to be strongest for modes of the same radial order, and for $\ell=0$ and $\ell=2$ p~modes in the asymptotic limit, p~modes of the same radial order differ in frequency by $\Delta \nu - \delta \nu \approx \Delta \nu$, where $\delta \nu$ is the small frequency spacing between $\ell=0$ and $\ell=2$ p~modes.
Once the tidal perturbation to the mode frequency becomes comparable to $\Delta \nu$, such coupling will be strong and will need to be accounted for. In our $\delta$\,Scuti model, this happens at orbital periods of roughly $P_{\rm orb} \lesssim 2 \, {\rm d}$ (see Figure \ref{fig:tidalsplitperechstar}). 

The same thing will happen for coupling between $\ell=1$ and $\ell=3$ modes. We do not expect this to affect observations of $\ell=1$ modes too much, because any $\ell=3$ component they obtain will be difficult to observe because of geometric cancellation \citep{dziembowski:77}. However, an $\ell=3$ mode that gains an $\ell=1$ component could be strongly altered because this component may dominate its observed brightness fluctuation. Similarly, $\ell=4$ modes mixing with $\ell=2$ modes may also be strongly altered, making them more visible in the star's power spectrum. 

For stronger tidal distortion, the $\ell_t=3$ component of the tidal distortion becomes more important, enabling coupling between modes differing by $\Delta \ell = 3$ and $\Delta \ell=1$. In this case, the star's eigenfunctions will be asymmetric across the $y$-$z$-plane, such that pulsations can be strongly trapped on either side of the star (i.e., the side facing toward or away from the companion). This leads to the ``tidal trapping" or ``single-sided pulsator" phenomenon seen in several systems (e.g., \citealt{handler:20,kurtz:20}). The likely signatures of strong tidal distortion and of coupling between modes of different $\ell$ is a more complex frequency spectrum than we predict. Examples include doublets with peaks of different amplitude, or with a central frequency component, or modes that produce more than three peaks in the power spectrum.

In general, tidal coupling could lead to complex behavior of the tidally perturbed pulsations, and the effects of strong tidal distortion, centrifugal, and the Coriolis force should all be taken into account. However, the simple case analyzed in this paper shows how pulsations can naturally align with the $x$, $y$, and $z-$axes of the star. More thorough calculations incorporating the effects listed above, and using larger networks of coupled modes, will be needed for a full understanding of pulsations of tidally distorted stars.

\subsection{Reanalysis of previous work}

Prior work on tidally tilted pulsators inferred mode geometries that should be revisited based on our results here. HD 74423 \citep{handler:20} and CO Cam \citep{kurtz:20} exhibit tidally trapped pulsations that are likely partially $Y_{10x}$ modes, but must also contain other components (e.g., radial modes and $Y_{22+}$ modes) in order to be tidally trapped. \cite{rappaport:21} and \cite{fuller:20} identified the tidally tilted pulsation in TIC 63328020 as an $m=\pm1$ mode propagating around the tidal axis. However, it now appears more likely that this mode is predominantly a $Y_{10y}$ mode, which nearly reproduces the observed amplitude and phase modulation. Similarly, \cite{kahraman:22} identified several nearly equal-amplitude doublets indicative of triaxial pulsations, with their inference of $m=0$ modes consistent with the $Y_{10x}$ modes we predict, while we interpret their claim of tidally tilted $|m|=1$ modes to instead be $Y_{10y}$ modes. Along the same lines, the $\ell=|m|=1$ and $\ell=|m|=2$ modes identified in TIC 68495594 by \cite{jayaraman:22} likely contain primarily $Y_{10y}$ and $Y_{22+}$ components, respectively, though both of these must have additional components in order to produce the observed amplitude and phase modulation.

\subsection{Tidally perturbed g~modes}

We concluded that g~modes are unlikely to be tidally tilted in synchronized binaries because the Coriolis force will be more important than the star's tidal distortion in affecting mode geometry. Yet \cite{vanreeth:22,vanreeth:23} have found several stars showing g~modes whose amplitudes are strongly modulated over the course of the orbit, such that they appear to be tidally tilted. However, an important feature of those pulsations is that their phase modulation is very small, in contrast to the phase jumps of $\approx$0.5 cycles typically observed in systems with tidally tilted p~modes. 

We believe the cause of the amplitude modulation may arise from ``tidal amplification" (see \citealt{fuller:20}) rather than tidal tilting. Since the structures of tidally distorted stars are no longer spherically symmetric, properties such as sound speed and Brunt-V\"ais\"al\"a frequency depend on tidal latitude. This could change the outer boundary of the g~mode cavity, such that a mode produces a larger flux perturbation on some parts of the star than others. In this scenario, the observed mode amplitude would vary throughout the orbit, but its phase would hardly change, consistent with the data. Our models would not capture this effect because they do not account for the gravity-darkened structure of the tidally distorted star. Future models should attempt to account for this effect.

\subsection{Non-synchronized and eccentric binaries}

Our analysis in this paper has been simplified by the assumption of a circular orbit with a tidally synchronized and aligned pulsator, such that the tidal distortion is static in the rotating frame of the star. If the star is not tidally synchronized, or if the orbit is eccentric, then the tidal distortion will not be static in the star's frame. It is unclear how the star's modes will behave in this case, and whether or not they will be tidally tilted. This should be studied in future work.

\cite{balona2018} showed that the p~modes observed in the heartbeat star KIC 4142768 have splittings by integer multiples of the orbital frequency, possibly indicative of amplitude modulation due to tidally tilted pulsations. However, the relative amplitude of the modes only varies by a small fraction in that case, whereas the triaxial pulsations shown in Figures \ref{fig:l1lightcurve}-\ref{fig:l2fourier} have order unity amplitude modulation. Another possibility is that there is non-linear coupling between p~modes and tidally excited g~modes that accounts for this splitting \citep{guo:19}, which may be observed in other eccentric p~mode pulsators \citep{hambleton:13,borkovits:14}. There is currently no clear-cut way to distinguish between these possibilities.

\section{Conclusions}

We demonstrated that tidally distorted stars tend to exhibit ``triaxial" pulsations that are aligned with one of the star's ellipsoidal axes. To do this, we used linear perturbation theory to compute the eigenfrequencies and eigenfunctions of a tidally coupled multiplet (i.e., the $2\ell+1$ different $m$ values of a mode with angular degree $\ell$ and radial order $n_{\rm pg}$). The $\ell_t=m_t=2$ component of the tidal distortion couples modes of different $m$, such that the eigenmodes are superpositions of multiple spherical harmonics $Y_{lm}$. When the tidal coupling terms are larger than the Coriolis perturbation terms, the modes become ``tidally tilted", and each mode has a nearly equal contribution of $Y_{lm}$ and $Y_{l-m}$.

The three tidally tilted dipole modes become $Y_{10x}$, $Y_{10y}$, and $Y_{10z}$ modes, each of which is symmetric about one of the three principal axes of the triaxial star (i.e., the tidal axis, the rotation axis, and the axis perpendicular to those two). The $Y_{10x}$ mode and $Y_{10y}$ mode exhibit amplitude modulation over the course of the orbit caused by the changing viewing angle between the line of sight and the tidal axis. Each mode thus creates equal-amplitude doublets spaced by twice the orbital frequency in the observed power spectrum. The $Y_{10z}$ mode is essentially the $\ell=1$, $m=0$ mode, whose amplitude is not modulated over the orbit, so it forms a singlet in the power spectrum.

The geometry of the five tidally tilted quadrupole modes depends on the mass ratio of the binary, but in many cases they are similar to spherical harmonics. Two of the modes have $Y_{21}$-like geometry, two have $Y_{22}$-like geometry, and the last mode is the normal $Y_{20}$ mode. In all cases, however, the modes are standing modes with a pattern that does not propagate around the $z$-axis. This means the observed mode amplitude and phase varies over the course of the orbit. The $Y_{21\pm}$ modes create doublets spaced by 2$\times$ the orbital frequency, the $Y_{22\pm}$ modes form doublets spaced by 4$\times$ the orbital frequency, and the $Y_{20z}$ mode forms a singlet. For quadrupole modes, the mode geometry depends on the mass ratio of the system, and there will often be some mixing between the $Y_{20z}$ modes and the $Y_{22\pm}$ modes.

We derived simple expressions for the frequency perturbations produced by tidal tilting (equations \ref{eq:omz}-\ref{eq:omx} for dipole p~modes), which is approximately the mode frequency times the tidal distortion amplitude. When these frequency perturbations become comparable to the large frequency spacing $\Delta \nu$, we expect tidal distortion to strongly couple modes of different $\ell$, not accounted for in this work (but see \citealt{fuller:20}). This will further distort the mode geometries, allowing for ``single-sided" pulsations confined to one side of the star, which we expect to occur in $\delta$\,Scuti binaries with periods less than $P_{\rm orb} \lesssim 2 \, {\rm d}$. Our model predicts p~modes to be tidally tilted in tidally circularized and synchronized binaries with $P_{\rm orb} \lesssim 3 \, {\rm d}$, and likely at longer periods up to $P \sim 6 \, {\rm d}$ depending on the mode frequency and spin rate of the star. We demonstrate that g~modes are unlikely to be tidally tilted in main sequence pulsators, remaining aligned with the star's spin axis, under most circumstances.

Although tidal tilting complicates the calculation of stellar oscillation modes, it also causes mode amplitude variations through the orbit which can be used to identify the mode geometry, allowing for mode identification. The $Y_{10x}$, $Y_{10y}$, and $Y_{10z}$ dipole modes can easily be distinguished, as already shown in \cite{zhang:24} and \cite{jayaraman:24}. Unfortunately, the $Y_{10z}$ mode may be difficult to distinguish from radial pulsation modes or $Y_{20z}$ modes since none of them exhibit orbital amplitude variation. Similarly, the $Y_{10x}$ and $Y_{10y}$ modes produce identical orbital variation to the quadrupolar $Y_{21-}$ and $Y_{21+}$ modes, so these could be difficult to distinguish. Nonetheless, mode identification enabled by tidal tilting may allow for detailed asteroseismic characterization of close binary stars in a manner not possible for single stars.

Along these lines, future observational work should search for more tidally tilted pulsators, whose clear signature is nearly equal-amplitude doublets spaced by twice the orbital frequency. These studies should use our results to identify the angular number $\ell$ and radial order $n_{\rm pg}$ of the observed tidally tilted modes to enable asteroseismic modeling. Future theoretical work should include tidal mode coupling between different angular number $\ell$, which will be especially important for $\ell=0$ and $\ell=2$ modes. It should also include coupling induced by the $\ell_t=3$ component of the tidal potential, which can cause the modes to be tidally trapped on one side of the star, and will be most important in very tidally distorted stars.

\section*{Acknowledgments}

GH is grateful for support by the Polish National Science Foundation (NCN) under grant nr.
2021/43/B/ST9/02972.


\bibliography{bib,CoreRotBib,CoreRotBib_2}

\begin{thebibliography}{}
\expandafter\ifx\csname natexlab\endcsname\relax\def\natexlab#1{#1}\fi
\providecommand{\url}[1]{\href{#1}{#1}}
\providecommand{\dodoi}[1]{doi:~\href{http://doi.org/#1}{\nolinkurl{#1}}}
\providecommand{\doeprint}[1]{\href{http://ascl.net/#1}{\nolinkurl{http://ascl.net/#1}}}
\providecommand{\doarXiv}[1]{\href{https://arxiv.org/abs/#1}{\nolinkurl{https://arxiv.org/abs/#1}}}

\bibitem[{{Balona}(2018)}]{balona2018}
{Balona}, L.~A. 2018, \mnras, 476, 4840, \dodoi{10.1093/mnras/sty544}

\bibitem[{{Bashi} {et~al.}(2023){Bashi}, {Mazeh}, \& {Faigler}}]{bashi:23}
{Bashi}, D., {Mazeh}, T., \& {Faigler}, S. 2023, \mnras, 522, 1184, \dodoi{10.1093/mnras/stad999}

\bibitem[{{Bildsten} {et~al.}(1996){Bildsten}, {Ushomirsky}, \& {Cutler}}]{bildsten:96}
{Bildsten}, L., {Ushomirsky}, G., \& {Cutler}, C. 1996, \apj, 460, 827, \dodoi{10.1086/177012}

\bibitem[{{Borkovits} {et~al.}(2014){Borkovits}, {Derekas}, {Fuller}, {Szab{\'o}}, {Pavlovski}, {Cs{\'a}k}, {D{\'o}zsa}, {Kov{\'a}cs}, {Szab{\'o}}, {Hambleton}, {Kinemuchi}, {Kolbas}, {Kurtz}, {Maloney}, {Pr{\v s}a}, {Southworth}, {Sztakovics}, {B{\'{\i}}r{\'o}}, \& {Jankovics}}]{borkovits:14}
{Borkovits}, T., {Derekas}, A., {Fuller}, J., {et~al.} 2014, \mnras, 443, 3068, \dodoi{10.1093/mnras/stu1379}

\bibitem[{{Bowman} {et~al.}(2019){Bowman}, {Johnston}, {Tkachenko}, {Mkrtichian}, {Gunsriwiwat}, \& {Aerts}}]{Bowman2019}
{Bowman}, D.~M., {Johnston}, C., {Tkachenko}, A., {et~al.} 2019, \apjl, 883, L26, \dodoi{10.3847/2041-8213/ab3fb2}

\bibitem[{{Chandrasekhar}(1963)}]{chandrasekhar1963}
{Chandrasekhar}, S. 1963, \apj, 138, 1182, \dodoi{10.1086/147716}

\bibitem[{{Chandrasekhar} \& {Lebovitz}(1963)}]{chandrasekhar1963b}
{Chandrasekhar}, S., \& {Lebovitz}, N.~R. 1963, \apj, 137, 1172, \dodoi{10.1086/147594}

\bibitem[{Dahlen \& Tromp(1998{\natexlab{a}})}]{dahlen:98}
Dahlen, F., \& Tromp, J. 1998{\natexlab{a}}, Theoretical Global Seismology (Princeton University Press).
\newblock \url{https://books.google.com/books?id=GWnuBws5gBEC}

\bibitem[{Dahlen \& Tromp(1998{\natexlab{b}})}]{dahlen1998}
---. 1998{\natexlab{b}}, Theoretical Global Seismology (Princeton University Press).
\newblock \url{https://books.google.com/books?id=GWnuBws5gBEC}

\bibitem[{{Denis}(1972)}]{denis1972}
{Denis}, J. 1972, \aap, 20, 151

\bibitem[{{Dziembowski}(1977)}]{dziembowski:77}
{Dziembowski}, W. 1977, \actaa, 27, 203

\bibitem[{{Fuller}(2014)}]{fuller:14}
{Fuller}, J. 2014, \icarus, 242, 283, \dodoi{10.1016/j.icarus.2014.08.006}

\bibitem[{{Fuller} {et~al.}(2020){Fuller}, {Kurtz}, {Handler}, \& {Rappaport}}]{fuller:20}
{Fuller}, J., {Kurtz}, D.~W., {Handler}, G., \& {Rappaport}, S. 2020, \mnras, 498, 5730, \dodoi{10.1093/mnras/staa2376}

\bibitem[{{Guo} {et~al.}(2019){Guo}, {Fuller}, {Shporer}, {Li}, {Hambleton}, {Manuel}, {Murphy}, \& {Isaacson}}]{guo:19}
{Guo}, Z., {Fuller}, J., {Shporer}, A., {et~al.} 2019, \apj, 885, 46, \dodoi{10.3847/1538-4357/ab41f6}

\bibitem[{{Hambleton} {et~al.}(2013){Hambleton}, {Kurtz}, {Pr{\v s}a}, {Guzik}, {Pavlovski}, {Bloemen}, {Southworth}, {Conroy}, {Littlefair}, \& {Fuller}}]{hambleton:13}
{Hambleton}, K.~M., {Kurtz}, D.~W., {Pr{\v s}a}, A., {et~al.} 2013, \mnras, 434, 925, \dodoi{10.1093/mnras/stt886}

\bibitem[{{Handler} {et~al.}(2020){Handler}, {Kurtz}, {Rappaport}, {Saio}, {Fuller}, {Jones}, {Guo}, {Chowdhury}, {Sowicka}, {Kahraman Ali{\c{c}}avu{\c{s}}}, {Streamer}, {Murphy}, {Gagliano}, {Jacobs}, \& {Vanderburg}}]{handler:20}
{Handler}, G., {Kurtz}, D.~W., {Rappaport}, S.~A., {et~al.} 2020, Nature Astronomy, 4, 684, \dodoi{10.1038/s41550-020-1035-1}

\bibitem[{{Holdsworth}(2021)}]{2021FrASS...8...31H}
{Holdsworth}, D.~L. 2021, Frontiers in Astronomy and Space Sciences, 8, 31, \dodoi{10.3389/fspas.2021.626398}

\bibitem[{{Jayaraman} {et~al.}(2022){Jayaraman}, {Handler}, {Rappaport}, {Fuller}, {Kurtz}, {Charpinet}, \& {Ricker}}]{jayaraman:22}
{Jayaraman}, R., {Handler}, G., {Rappaport}, S.~A., {et~al.} 2022, \apjl, 928, L14, \dodoi{10.3847/2041-8213/ac5c59}

\bibitem[{Jayaraman {et~al.}(2024)Jayaraman, Rappaport, Powell, Handler, Omohundro, Gagliano, Kostov, Fuller, Kurtz, Zhang, \& Ricker}]{jayaraman:24}
Jayaraman, R., Rappaport, S., Powell, B., {et~al.} 2024, TIC 435850195: The Second Tri-Axial, Tidally Tilted Pulsator.
\newblock \doarXiv{2409.03815}

\bibitem[{{Jennings} {et~al.}(2024){Jennings}, {Southworth}, {Pavlovski}, \& {Van Reeth}}]{jennings:24}
{Jennings}, Z., {Southworth}, J., {Pavlovski}, K., \& {Van Reeth}, T. 2024, \mnras, 527, 4052, \dodoi{10.1093/mnras/stad3427}

\bibitem[{{Johnston} {et~al.}(2023){Johnston}, {Tkachenko}, {Van Reeth}, {Bowman}, {Pavlovski}, {Sana}, \& {Sekaran}}]{johnston:23}
{Johnston}, C., {Tkachenko}, A., {Van Reeth}, T., {et~al.} 2023, \aap, 670, A167, \dodoi{10.1051/0004-6361/202244808}

\bibitem[{{Kahraman Ali{\c{c}}avu{\c{s}}} {et~al.}(2022){Kahraman Ali{\c{c}}avu{\c{s}}}, {Handler}, {Ali{\c{c}}avu{\c{s}}}, {De Cat}, {Bedding}, {Lampens}, {Ekinci}, {G{\"u}m{\"u}{\textcommabelow s}}, \& {Leone}}]{kahraman:22}
{Kahraman Ali{\c{c}}avu{\c{s}}}, F., {Handler}, G., {Ali{\c{c}}avu{\c{s}}}, F., {et~al.} 2022, \mnras, 510, 1413, \dodoi{10.1093/mnras/stab3515}

\bibitem[{{Kurtz}(1982)}]{1982MNRAS.200..807K}
{Kurtz}, D.~W. 1982, \mnras, 200, 807, \dodoi{10.1093/mnras/200.3.807}

\bibitem[{{Kurtz}(2022)}]{2022ARA&A..60...31K}
---. 2022, \araa, 60, 31, \dodoi{10.1146/annurev-astro-052920-094232}

\bibitem[{{Kurtz} \& {Shibahashi}(1986)}]{kurtz:86}
{Kurtz}, D.~W., \& {Shibahashi}, H. 1986, \mnras, 223, 557, \dodoi{10.1093/mnras/223.3.557}

\bibitem[{{Kurtz} {et~al.}(2020){Kurtz}, {Handler}, {Rappaport}, {Saio}, {Fuller}, {Jacobs}, {Schmitt}, {Jones}, {Vanderburg}, {LaCourse}, {Nelson}, {Kahraman Ali{\c{c}}avu{\c{s}}}, \& {Giarrusso}}]{kurtz:20}
{Kurtz}, D.~W., {Handler}, G., {Rappaport}, S.~A., {et~al.} 2020, \mnras, 494, 5118, \dodoi{10.1093/mnras/staa989}

\bibitem[{{Lee} \& {Saio}(1997)}]{lee:97}
{Lee}, U., \& {Saio}, H. 1997, \apj, 491, 839

\bibitem[{{Martens} \& {Smeyers}(1982)}]{martens1982}
{Martens}, L., \& {Smeyers}, P. 1982, \aap, 106, 317

\bibitem[{{Martens} \& {Smeyers}(1986)}]{martens1986}
---. 1986, \aap, 155, 211

\bibitem[{{Paxton} {et~al.}(2011){Paxton}, {Bildsten}, {Dotter}, {Herwig}, {Lesaffre}, \& {Timmes}}]{paxton:11}
{Paxton}, B., {Bildsten}, L., {Dotter}, A., {et~al.} 2011, \apjs, 192, 3, \dodoi{10.1088/0067-0049/192/1/3}

\bibitem[{{Paxton} {et~al.}(2013){Paxton}, {Cantiello}, {Arras}, {Bildsten}, {Brown}, {Dotter}, {Mankovich}, {Montgomery}, {Stello}, {Timmes}, \& {Townsend}}]{paxton:13}
{Paxton}, B., {Cantiello}, M., {Arras}, P., {et~al.} 2013, \apjs, 208, 4, \dodoi{10.1088/0067-0049/208/1/4}

\bibitem[{{Paxton} {et~al.}(2015){Paxton}, {Marchant}, {Schwab}, {Bauer}, {Bildsten}, {Cantiello}, {Dessart}, {Farmer}, {Hu}, {Langer}, {Townsend}, {Townsley}, \& {Timmes}}]{paxton:15}
{Paxton}, B., {Marchant}, P., {Schwab}, J., {et~al.} 2015, \apjs, 220, 15, \dodoi{10.1088/0067-0049/220/1/15}

\bibitem[{{Paxton} {et~al.}(2018){Paxton}, {Schwab}, {Bauer}, {Bildsten}, {Blinnikov}, {Duffell}, {Farmer}, {Goldberg}, {Marchant}, {Sorokina}, {Thoul}, {Townsend}, \& {Timmes}}]{paxton:18}
{Paxton}, B., {Schwab}, J., {Bauer}, E.~B., {et~al.} 2018, \apjs, 234, 34, \dodoi{10.3847/1538-4365/aaa5a8}

\bibitem[{{Paxton} {et~al.}(2019){Paxton}, {Smolec}, {Gautschy}, {Bildsten}, {Cantiello}, {Dotter}, {Farmer}, {Goldberg}, {Jermyn}, {Kanbur}, {Marchant}, {Schwab}, {Thoul}, {Townsend}, {Wolf}, {Zhang}, \& {Timmes}}]{paxton:19}
{Paxton}, B., {Smolec}, R., {Gautschy}, A., {et~al.} 2019, arXiv e-prints.
\newblock \doarXiv{1903.01426}

\bibitem[{{Rappaport} {et~al.}(2021){Rappaport}, {Kurtz}, {Handler}, {Jones}, {Nelson}, {Saio}, {Fuller}, {Holdsworth}, {Vanderburg}, {{\v{Z}}{\'a}k}, {Skarka}, {Aiken}, {Maxted}, {Stevens}, {Feliz}, \& {Kahraman Ali{\c{c}}avu{\c{s}}}}]{rappaport:21}
{Rappaport}, S.~A., {Kurtz}, D.~W., {Handler}, G., {et~al.} 2021, \mnras, 503, 254, \dodoi{10.1093/mnras/stab336}

\bibitem[{{Reed} {et~al.}(2005){Reed}, {Brondel}, \& {Kawaler}}]{reed:05}
{Reed}, M.~D., {Brondel}, B.~J., \& {Kawaler}, S.~D. 2005, \apj, 634, 602, \dodoi{10.1086/491666}

\bibitem[{{Reed} {et~al.}(2011){Reed}, {Harms}, {Poindexter}, {Zhou}, {Eggen}, {Morris}, {Quint}, {McDaniel}, {Baran}, {Dolez}, {Kawaler}, {Kurtz}, {Moskalik}, {Riddle}, {Zola}, {{\O}stensen}, {Solheim}, {Kepler}, {Costa}, {Provencal}, {Mullally}, {Winget}, {Vuckovic}, {Crowe}, {Terry}, {Avila}, {Berkey}, {Stewart}, {Bodnarik}, {Bolton}, {Binder}, {Sekiguchi}, {Sullivan}, {Kim}, {Chen}, {Chen}, {Lin}, {Jian}, {Wu}, {Gou}, {Liu}, {Leibowitz}, {Lipkin}, {Akan}, {Cakirli}, {Janulis}, {Pretorius}, {Ogloza}, {Stachowski}, {Paparo}, {Szabo}, {Csubry}, {Zsuffa}, {Silvotti}, {Marinoni}, {Bruni}, {Vauclair}, {Chevreton}, {Matthews}, {Cameron}, \& {Pablo}}]{reed:11}
{Reed}, M.~D., {Harms}, S.~L., {Poindexter}, S., {et~al.} 2011, \mnras, 412, 371, \dodoi{10.1111/j.1365-2966.2010.17912.x}

\bibitem[{{Reyniers} \& {Smeyers}(2003{\natexlab{a}})}]{reyniers2003}
{Reyniers}, K., \& {Smeyers}, P. 2003{\natexlab{a}}, \aap, 404, 1051, \dodoi{10.1051/0004-6361:20030501}

\bibitem[{{Reyniers} \& {Smeyers}(2003{\natexlab{b}})}]{reyniers2003b}
---. 2003{\natexlab{b}}, \aap, 409, 677, \dodoi{10.1051/0004-6361:20031098}

\bibitem[{{Saio}(1981)}]{saio:81}
{Saio}, H. 1981, \apj, 244, 299, \dodoi{10.1086/158708}

\bibitem[{{Smeyers} \& {Martens}(1983)}]{smeyers1983}
{Smeyers}, P., \& {Martens}, L. 1983, \aap, 125, 193

\bibitem[{{Tassoul} \& {Tassoul}(1967)}]{tassoul1967}
{Tassoul}, M., \& {Tassoul}, J.~L. 1967, \apj, 150, 213, \dodoi{10.1086/149322}

\bibitem[{{Townsend} \& {Teitler}(2013)}]{townsend:13}
{Townsend}, R.~H.~D., \& {Teitler}, S.~A. 2013, \mnras, 435, 3406, \dodoi{10.1093/mnras/stt1533}

\bibitem[{{Van Reeth} {et~al.}(2023){Van Reeth}, {Johnston}, {Southworth}, {Fuller}, {Bowman}, {Poniatowski}, \& {Van Beeck}}]{vanreeth:23}
{Van Reeth}, T., {Johnston}, C., {Southworth}, J., {et~al.} 2023, \aap, 671, A121, \dodoi{10.1051/0004-6361/202245460}

\bibitem[{{Van Reeth} {et~al.}(2022){Van Reeth}, {Southworth}, {Van Beeck}, \& {Bowman}}]{vanreeth:22}
{Van Reeth}, T., {Southworth}, J., {Van Beeck}, J., \& {Bowman}, D.~M. 2022, \aap, 659, A177, \dodoi{10.1051/0004-6361/202142833}

\bibitem[{{Zhang} {et~al.}(2024){Zhang}, {Rappaport}, {Jayaraman}, {Kurtz}, {Handler}, {Fuller}, \& {Borkovits}}]{zhang:24}
{Zhang}, V., {Rappaport}, S., {Jayaraman}, R., {et~al.} 2024, \mnras, 528, 3378, \dodoi{10.1093/mnras/stae010}

\end{thebibliography}

\newpage
\appendix

\section{Tidal Coupling Integrals}

Here we provide the form of the kinetic and potential energy overlap integrals from Appendix D of \cite{dahlen1998}. That work was applied to the centrifugal distortion (which has $\ell_t=2$, $m_t=0$). We rearrange their expressions, simplifying them to the case of $\ell=\ell'$. However, we generalize them to apply to both the centrifugal distortion and the tidal distortion, which has components with $\ell_t=2$, $m_t=0$ and $m_t=\pm 2$. The kinetic energy overlap has form
\beq
\label{dT}
\delta T_{\alpha \alpha'} \propto X_{\ell \ell' \ell_t}^{m m' m_t} \int^R_0 \frac{2}{3} \epsilon \rho r^2 \big[ \bar{T}_\rho - (\eta + 3) \check{T}_\rho \big] dr 
\eeq
with 
\beq
\label{dT2}
\bar{T}_\rho = -Y_{\ell_t} U V' - Y_{\ell_t} U' V 
\eeq
and
\beq
\label{dT3}
\check{T}_\rho =  U U' + Z_{\ell \ell' \ell_t } V V' \, .
\eeq
In these expressions, $U$ is the radial displacement associated with mode $\alpha$, and $U'$ is the radial displacement for mode $\alpha'$. Similarly, $V$ is the horizontal displacement. Also,
\begin{equation}
\label{y}
Y_{\ell_t} = \frac{\ell_t(\ell_t+1)}{2} \, ,
\end{equation}
\begin{equation}
\label{z}
Z_{\ell \ell' \ell_t } = \frac{1}{2} \big[ \ell(\ell+1) + \ell'(\ell'+1) - \ell_t(\ell_t+1) \big] \, .
\end{equation}
We have already factored out the time and angular dependence, i.e., the full displacement is $\bxi = U Y_{\ell m} e^{-i \omega t} + V r\nabla_\perp Y_{\ell m} e^{-i \omega t}$.

The potential energy coupling terms have form
\beq
\label{dV}
\delta V_{\alpha \alpha'} \propto X_{\ell \ell' \ell_t}^{m m' m_t} \int^R_0 \frac{2}{3} \epsilon r^2 \bigg( \kappa \big[\bar{V}_\kappa - (\eta +1) \check{V}_\kappa \big] + \rho \big[ \bar{V}_\rho - (\eta + 3) \check{V}_\rho \big] \bigg) dr 
\eeq
and the incompressibility is $\kappa = \rho c_s^2 = \Gamma_1 p$, where $c_s$ is the sound speed, $p$ is the pressure and $\Gamma_1$ is the adiabatic index. The integrand components are 
\begin{align}
\bar{V}_\kappa &= - \partial U/\partial r (\partial U'/\partial r + f') - \partial U'/\partial r (\partial U/\partial r + f) \nonumber \\
& -Y_{\ell_t} V (\partial U'/\partial r + f')/r - Y_{\ell_t} V' (\partial U/\partial r + f )/r
\end{align}
\begin{align}
\check{V}_\kappa &= \frac{1}{2} (-\partial U/\partial r +f)(\partial U'/\partial r + f') + \frac{1}{2} (-\partial U'/\partial r + f')(\partial U/\partial r + f) \nonumber \\
& + Y_{\ell_t} V (\partial U'/\partial r + f')/r + Y_{\ell_t} V' (\partial U/\partial r + f)/r
\end{align}
\begin{align}
\bar{V}_\rho &= (r \partial P/\partial r + 4 \pi G \rho r U + g U) f' + (r \partial P'/\partial r + 4 \pi G \rho r U' + g U') f \nonumber \\
& - Y_{\ell_t} g V U'/r - Y_{\ell_t} g V' U/r + 3 g U U'/r + 3 g U' U/r \nonumber \\
& + Z_{\ell \ell' \ell_t } P V'/r + Z_{\ell \ell' \ell_t} P' V/r \nonumber \\
&- \ell(\ell+1) P U'/r - \ell'(\ell'+1) P' U/r 
\end{align}
\begin{align}
\check{V}_\rho &= U \partial P'/\partial r + U' \partial P/\partial r + 4 \pi G \rho U U' + 4 \pi G \rho U' U \nonumber \\
& - Y_{\ell_t} g U V'/r - Y_{\ell_t} g U' V/r + Z_{\ell \ell' \ell_t } V P'/r + Z_{\ell \ell' \ell_t } V'P/r 
\end{align}
Here, $P$ is the the Eulerian gravitational potential perturbation $\delta \Phi = P Y_{\ell m} e^{- i \omega t}$, and we define $f = \big[2 U - \ell (\ell+1) V \big]/r$.

\section{Summary of Tidally Tilted Modes}
\label{sec:appendix}

Here we decompose triaxial modes into their spherical harmonic components in compact matrix notation. For $\ell = 1$ modes:

$$
\begin{bmatrix}
Y_{10x} \\
Y_{10y} \\
Y_{10z} 
\end{bmatrix}
= \frac{1}{\sqrt{2}}
\begin{bmatrix}
1 & 0 & -1 \\
1 & 0 & 1 \\
0 & \sqrt{2} & 0 
\end{bmatrix}
\begin{bmatrix}
Y_{11} \\ Y_{10} \\ Y_{1-1}
\end{bmatrix}
$$
\vspace{10pt}

\noindent
These modes are appropriately normalized in the same manner as spherical harmonics, e.g., 
\begin{equation}
\int dS \, Y_{10x}^2 = 1 \, ,
\end{equation}
where the integral is taken over a spherical surface. Note that all of the triaxial modes are purely real or imaginary functions because they contain equal components of $+m$ and $-m$ spherical harmonics.

\vspace{1cm}
For $\ell = 2$ modes:
$$
\begin{bmatrix}
Y_{21-} \\
Y_{21+} \\
Y_{22-} \\
Y_{22+} \\
Y_{20z}
\end{bmatrix}
= \frac{1}{\sqrt{2}}
\begin{bmatrix}
0 & 1 & 0 & -1 & 0\\
0 & 1 & 0 & 1 & 0 \\
1 & 0 & 0 & 0 & -1 \\
1 & 0 & 0 & 0 & 1 \\
0 & 0 & \sqrt{2} & 0 & 0
\end{bmatrix}
\begin{bmatrix}
Y_{22} \\ Y_{21} \\ Y_{20} \\ Y_{2-1} \\ Y_{2-2}
\end{bmatrix}
$$

For $\ell = 3$ modes:
$$
\begin{bmatrix}
Y_{31-} \\
Y_{31+} \\
Y_{33-} \\
Y_{33+} \\
Y_{32-} \\
Y_{32+} \\
Y_{30z}
\end{bmatrix}
= \frac{1}{\sqrt{2}}
\begin{bmatrix}
0 & 0 & 1 & 0 & -1 & 0 & 0\\
0 & 0 & 1 & 0 & 1 & 0 & 0 \\
1 & 0 & 0 & 0 & 0 & 0 & -1 \\
1 & 0 & 0 & 0 & 0 & 0 & 1 \\
0 & 1 & 0 & 0 & 0 & -1 & 0 \\
0 & 1 & 0 & 0 & 0 & 1 & 0 \\
0 & 0 & 0 & \sqrt{2} & 0 & 0 & 0
\end{bmatrix}
\begin{bmatrix}
Y_{33} \\ Y_{32} \\ Y_{31} \\ Y_{30} \\ Y_{3-1} \\ Y_{3-2} \\ Y_{3-3}
\end{bmatrix}
$$
These results for $\ell=2$ and $\ell=3$ modes hold only if the off-diagonal coupling terms are small compared to the difference between the diagonal self-coupling terms.  This is, in fact, the case as long as the mass ratio in the binary ($M_c/M$) is somewhat less than $\sim$unity.

\vspace{10pt}


\end{document}